# Molecular ink-based synthesis of Bi(S$_z$Se$_{1-z}$)(I$_x$Br$_{1-x}$) solid solutions as tuneable materials for sustainable energy applications

D. Rovira[1,2,*], I. Caño[1,2], C. López[2,3], A. Navarro-Güell[1,2], J.M. Asensi[4,5], L. Calvo-Barrio[6,7], L. Garcia-Carreras[6], X. Alcobe[6], L. Cerqueira[8], V. Corregidor[8], Y. Sanchez[9], S. Lanzalaco[2,12], A. Jimenez-Arguijo[1,2], O. El Khouja[1,2], J. W. Turnley[10,11], R. Agrawal[10], C. Cazorla[2,3], J. Puigdollers[1,2], E. Saucedo[1,2,+]

[1]Universitat Politècnica de Catalunya (UPC), Photovoltaic Lab – Micro and Nano Technologies Group (MNT), Electronic Engineering Department, EEBE, Av Eduard Maristany 10-14, Barcelona 08019, Spain
[2]Universitat Politècnica de Catalunya (UPC), Barcelona Centre for Multiscale Science & Engineering, Av Eduard Maristany 10-14, Barcelona 08019, Spain
[3]Universitat Politècnica de Catalunya (UPC), Group of Characterization of Materials (GCM), Physics Department, EEBE, Av Eduard Maristany 10-14, Barcelona 08019, Spain
[4]Universitat de Barcelona (UB), Applied Physics Department, C. Martí I Franquès 1, Barcelona 08028, Spain
[5]Institute of Nanoscience and Nanotechnology (IN2UB), Universitat de Barcelona, 08028 Barcelona, Spain.
[6]Universitat de Barcelona (UB), Scientific and Technological Centers (CCiTUB), C. Lluís Solé i Sabaris 1-3, Barcelona 08028, Spain
[7]Universitat de Barcelona (UB), Electronics and Biomedical Engineering Department, C. Martí I Franquès 1, Barcelona 08028, Spain
[8]Centro de Ciências E Tecnologias Nucleares (C2TN), Instituto Superior Técnico, Universidade de Lisboa, E.N. 10 Ao Km 139.7, 2695-066, Bobadela LRS, Portugal
[9]Catalonia Institute for Energy Research – IREC, Jardins de les Dones de Negre 1, Sant Adrià de Besòs 08930, Spain
[10]Davidson School of Chemical Engineering, Purdue University, West Lafayette, Indiana 47907, USA
[11]Department of Chemical Engineering and Materials Science, Michigan State University, East Lansing, Michigan 48824, USA
[12]Universitat Politècnica de Catalunya, Innovation in Materials and Molecular Engineering (IMEM), Department of Chemical Engineering, Av Eduard Maristany 10-14, Barcelona 08019, Spain

Corresponding authors
[*]david.rovira.ferrer@upc.edu
[+]edgardo.saucedo@upc.edu

Quasi-one-dimensional (Q-1D) van der Waals chalcohalides have emerged as promising materials for advanced energy applications, combining tunable optoelectronic properties and composed by earth-abundant and non-toxic elements. However, their widespread application remains hindered by challenges such as anisotropic crystal growth, composition control and lack of knowledge on optoelectronic properties. A deeper understanding of the intrinsic limitations of these materials, as well as viable defect mitigation strategies like the engineering of solid solutions, is critical. This work presents a low-temperature synthesis route based on molecular ink deposition enabling direct crystallization of tunable Bi(S$_z$Se$_{1-z}$)(I$_x$Br$_{1-x}$) solid solutions without need for binary chalcogenide precursors. This approach yields phase-pure films with precise control over morphology, composition, and crystallographic orientation. XRD analysis and DFT calculations confirm the formation of homogeneous solid solutions, while optoelectronic measurements reveal the distinct roles of halogen and chalcogen anions in tuning bandgap energy and carrier type, with Se shifting downwards the conduction band. The versatility of this synthesis technique enables morphology control ranging from compact films to rod-shaped microcrystals, expanding the functional adaptability of these materials. These findings offer a foundational framework for defect engineering and the scalable integration of chalcohalides in next-generation energy technologies, including photovoltaics, photocatalysis, thermoelectrics, and chemical sensing.

**Keywords:**

Chalcohalides, Q-1D materials. Molecular inks, optoelectronic tuneability, solid solutions.

## 1. Introduction

Two-dimensional (2D) van der Waals (vdW) materials, such as $MoS_2$ and graphene, have garnered significant attention for functional applications due to their high surface area, excellent carrier mobility, and mechanical robustness.[1–5] However, their limited tunability in terms of bandgap and electronic properties restricts their versatility, highlighting the need for alternative materials with greater compositional and structural adaptability. In this context, quasi-one-dimensional (Q-1D) vdW materials, such as MChX (M = Sb, Bi; Ch = S, Se; X = I, Br) chalcohalides, are emerging as promising candidates for next-generation functional applications.[6,7] Chalcohalide materials crystallize in an orthorhombic crystalline structure with symmetry characterized by the Pnma space group and consisting of $(M_2Ch_2X_2)_n$ covalently bonded ribbons extending along one crystallographic axis, with weak vdW interactions binding the ribbons in orthogonal directions. The Q-1D nature originates from the $ns^2$ lone pair of pnictogen cations, which distorts their coordination environment and creates an asymmetric electronic density.[8] This structure has been proposed to lead to perovskite-like electronic characteristics like defect tolerance, high dielectric constants, and efficient charge transport resulting in unique, anisotropic optoelectronic properties. Their intrinsic anisotropy in transport and optical properties makes them promise for directional control in electronic and photonic systems. The strong quantum confinement and edge-dominated surfaces enhance their tunability and chemical reactivity, offering a wide bandgap range, from 1 to 3.5 eV.[9–12] Importantly, the presence of heavy elements in many MChX compounds enhances phonon scattering, making them suitable for thermoelectric applications.[7,13,14]

Despite their promising optoelectronic properties, the current performance of chalcohalide-based optoelectronic devices remains significantly poor.[15–17] These results underscore the urgent need for further investigation into the fundamental limitations of chalcohalide semiconductors and the development of strategies to overcome them. One key challenge arises from their intrinsic anisotropy. While enabling high carrier mobility along covalent directions, it promotes ribbon-like crystallite growth, hindering the formation of compact, uniform films and leading to poor in-plane conductivity and inefficient charge extraction. Advancing morphology control techniques for Q-1D materials is crucial to unlocking their full potential in energy applications. Also, although chalcohalides have been suggested to exhibit defect tolerance, recent experimental and theoretical studies[18,19] suggest that deep defect states still play a detrimental role. Among the strategies proposed to mitigate these effects, the use of solid solutions has shown promise due to ionic radius mismatch suppressing ion exchangeability and increasing the formation energy of point defects.[19]

Various physical and chemical synthesis routes for Bi-based chalcohalides have been established in the literature. They include hydrothermal deposition, co-evaporation of precursor binary chalcogenides followed by reactive annealing in a halogen-rich atmosphere, or sonochemical methods using elemental compounds.[20–25] However, these methods are often limited to one or two materials within the chalcohalide family. The use of molecular inks has also been reported,[16,17] and this method has proven its effectiveness in the synthesis of other chalcogenides. "Alkahest" systems based on hydrazine or amine-thiols, as well as metal-organic-chalcogen complex-based systems, have reached a high level of maturity in producing high-efficiency devices.[25–30] Chalcogenourea ligands (thioureas/selenoureas) are powerful enablers for low-temperature conversion of molecular inks into metal chalcogenides. They combine facile chalcogen transfer chemistry, tunable coordination to metal centers, and decomposition kinetics that can be tailored by ligand substitution.[31,32] Seminal work from the Owen group showed that substituted thioureas act as single-source or co-precursors whose reactivity spans many orders of magnitude.[33] By choosing ligand substituents one can control the conversion rate, nucleation density and ultimately crystal size and composition during mild thermolysis. Recent studies have extended this concept to selenoureas and other chalcogen-rich precursors, demonstrating low-temperature formation of metal selenides and solution-processed chalcogenide thin films from molecular formulations.[34]

Building on this knowledge, we develop a general chemical ink for synthesizing chalcohalides. We synthetize for the first time the complete Bi-chalcohalide family, including BiSI, BiSBr, BiSeI, and BiSeBr, and their corresponding solid solutions, using the same solvent system, with precise control over morphology, stoichiometry, and optoelectronic properties. The approach is based on the spin-coating of precursor inks dissolved in *N,N*-dimethylformamide (DMF), followed by a low temperature thermal annealing step to drive crystallization of the ternary phase (**Fig. 1**). The synthesis and characterization of Bi-based chalcohalide solid solutions have been explored as a foundation for future material engineering strategies. In the first part of the study, we show the phase purity of the synthesized compounds with no segregation of secondary phases. In-depth investigation of their structural characteristics is performed through X-ray diffraction analysis. We demonstrate excellent compositional control, supporting the viability of this direct synthesis method without the need for binary chalcogenide precursor layers.[35] Also, we develop strategies that facilitate control of orientation and film morphology, crucial for functionality. In the second part, we analyze the optoelectronic properties of solid solutions, disentangling the individual roles of the chalcogen and halogen anions in tuning charge carrier type, and optical absorption, highlighting their potential for band structure engineering. Density Functional Theory (DFT) calculations were conducted to complement the analysis of the optoelectronic and structural properties of solid solutions. This combination of tunable material chemistry and scalable processing places chalcohalide semiconductors as promising candidates for a wide range of applications including photovoltaics, photocatalysis, thermoelectricity, and sensing.

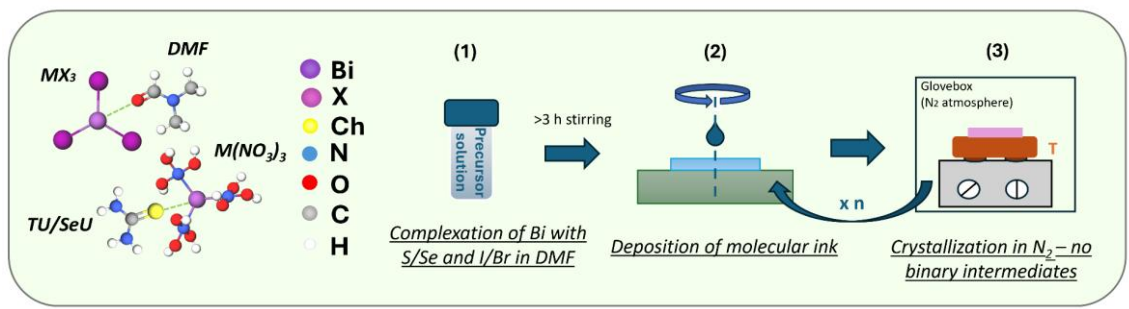

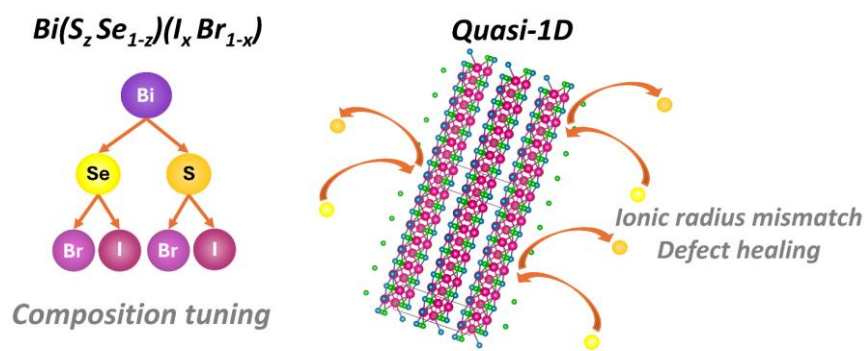

**Fig. 1.** The solution-processed synthesis enables direct formation of Bi-based chalcohalides with controlled composition and crystallinity. The resulting solid solutions offer both optoelectronic tunability and intrinsic point-defect healing through lattice strain relaxation

## 2. Results and discussion
### 2.1. Solid solution phase verification

The characteristic X-ray diffraction (XRD) patterns for the complete set of Bi-chalcohalide compounds listed in **Table S1** are shown in **Fig. 2**. A magnified view of the peak corresponding to the (110) plane is included to highlight the shift in its position for the solid solutions. Quantitative pattern matching refinement, performed using the Le Bail method, confirmed an orthorhombic structure with a Pnma space group for all compounds.[36] The samples analysed were synthesized on fluorine doped tin oxide (FTO) coated glass ($SnO_2$, tetragonal structure with $P4_2/mnm$ space group), whose diffraction peaks are marked with grey dashed lines. For the halogen solid solutions $BiSI_{0.5}Br_{0.5}$ and $BiSI_{0.7}Br_{0.3}$, a minor bromine-poor phase, $Bi_{12.89}Br_{2.85}S_{17.15}$, was detected. The refined fittings and variation of cell parameters are provided in the **Supplementary Information (SI)**.

**Fig. 2d and 2f** clearly illustrate the shift in the position of the (110) peak. The gradual substitution of sulphur with selenium in BiSBr results in a shift toward lower angles, indicating an expansion of the unit cell due to the larger ionic radius of selenium. Conversely, when iodine is replaced with bromine in BiSI, the unit cell contracts, causing the diffraction peaks to shift toward higher angles. This behaviour is further elucidated by the refined lattice parameters of the final structures.

For the four parent compounds (BiSI, BiSBr, BiSeI, and BiSeBr), the obtained lattice parameters were $a$ = 8.45(12) ; 8.10(29) ; 8.64(71) ; 8.21(41) Å, $b$ = 10.11(92) ; 9.77(73) ; 10.49(28) ; 10.44(73) Å, and $c$ = 4.17(13) ; 4.04(39) ; 4.19(91) ; 4.11(05) Å, respectively. These values are in excellent agreement with previously reported data.[35,37–40]

When inspecting the cell parameters of the solid solutions (illustrated in **Fig. S1** and **Fig. S2**), the lattice expansion or contraction resulting from ion substitution can be quantitatively observed. Substituting ions with smaller radii for those with larger radii leads to an increase in the unit cell volume. This trend is typically described through Vegard's rule which states that the lattice parameter of an alloy varies linearly with the concentration of its constituent elements.[41] Both chalcogen and halogen solid solutions appears to follow reasonably well Vegard's Rule (see **Fig. 2e and 2g**). This supports the formation of a continuous solid solution, a structural stability across the range and the systematic incorporation of the substituent into the lattice.

In the chalcogen solid solution, a clear linear increase in all lattice parameters is observed as sulphur is progressively replaced by selenium. Particularly significant is the expansion of the *b*-parameter, which increases by 6.86%, from 9.77 Å in BiSBr to 10.44 Å in BiSeBr. In contrast, for the *a*- and *c*-parameters the increase is much lower, 1.37% and 1.64% respectively. This fact strongly confirms that the b-direction corresponds to the one were Bi-Ch interactions are predominant.

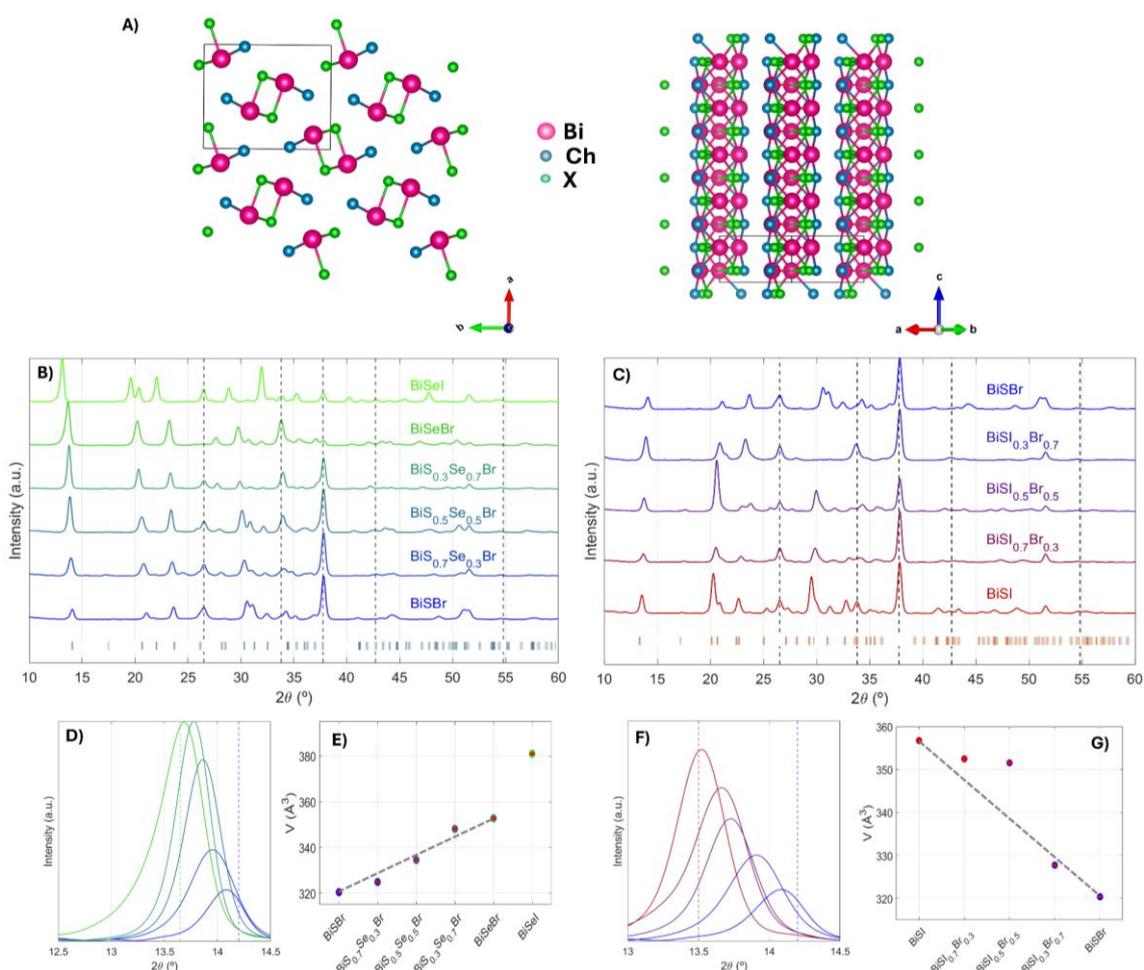

**Fig. 2. (a)** Crystal structure of MChX **(b,c)** XRD patterns for the Bi-chalcohalide solid solutions and parent compounds, with the reference (hkl) peaks of BiSBr and BiSI highlighted in blue and red, respecitevely. FTO peaks are indicated as grey dashed lines **(d,f)** Magnified views of the peak corresponding to the (110) plane of the chalcohalides. **(e, g)** Unit cell volume for Bi-chalcohalide solid solutions. The dashed line indicates the theoretical evolution following Vegard's Rule (error bars shown in red).

In contrast, the halogen solid solution does not exhibit a strictly linear decrease in lattice parameters upon substituting iodine with bromine; instead, a negative bowing effect is observed. (**Fig. S2**). A plateau in the lattice values is observed until bromine becomes the dominant halide (x ≥ 0.5), beyond which the parameters drop sharply toward the values of BiSBr. Additionally, the relative variation in lattice dimensions is slightly greater along the *a*-axis (4.28 %) than along the *b*- and *c*-axes (3.37 % and 3.05 %, respectively), differently to the trend observed in the chalcogen series. This is an agreement with the fact that the Bi-X interactions are mostly placed in the *a*-direction.

These results highlight the distinct local bonding effects that define the structural behaviour of these materials, which consist of Q-1D motifs. The structure combines covalent bonding within the 1D ribbons (along the *c*-axis) and vdW interactions between the ribbons in the *ab*-plane. Chalcogen and pnictogen ions constitute the core of the 1D ribbons, while halogen ions are located on the edges. Halogen substitution influence interchain interactions predominantly through halogen-halogen vdW interactions and does not strictly follow Vegard's law. Unlike strong covalent bonding, vdW interactions rely on the presence of transient dipoles. The easier it is to distort the electron cloud of an ion, the larger the induced dipole moment and, consequently, the stronger the dispersion forces. These interactions are therefore strongly governed by the polarizability of the constituent ions. These vdW interactions are prominently affected by the polarizability of the ions involved. The higher polarizability of I$^-$ may provide stronger and stiffer interactions, especially in the ribbon-stacking direction. When iodine is the dominant halogen, this stronger inter-ribbon locks the lattice parameters, having a rigid structure. As bromine content increases, the weaker vdW interactions that introduces thanks to its lower polarizability possibly relax these constraints, allowing lattice parameters to adjust more freely. Then, a sharp drop in lattice constants is observed when bromine becomes the predominant halogen. Consequently, halogen substitution is expected to play a key role in determining morphology and stacking stability of the Q-1D structures in these materials (see **Morphology Control Section**).

To further support this hypothesis, ab initio calculations based on DFT were carried out for the four parent compounds to quantify the strength of vdW interactions (see **Experimental Section** for more details). The results are summarized in **Table 1**. The computed lattice parameters, included in **Table 1**, show good agreement with the experimental values, with a slight underestimation. This consistency serves as a validation of the reliability of the performed simulations. Importantly, iodine-based compounds display higher normalized vdW energy differences compared to their bromine counterparts (BiSeI > BiSI > BiSeBr > BiSBr). In particular, the highest vdW interaction energy (i.e., the difference between total ground energies calculated with and without dispersion interactions) is obtained for BiSeI, with 8.19 meV/Å$^3$, significantly higher than the 6.25 meV/Å$^3$ of BiSBr. Analyzing the impact of halogen and chalcogen substitutions reveals distinct trends: replacing S with Se in BiSI and BiSBr increases the vdW interaction energy by 10.8% and 12.8%, respectively, while substituting Br with I in BiSBr and BiSeBr results in larger increases of 18.2% and 16.2%. These results highlight the stronger influence of halogen substitution—particularly iodine—on enhancing vdW interactions. These theoretical results support the general wisdom that enhanced inter-ribbon interactions are formed with iodine, contributing to greater structural rigidity and anisotropy. The higher polarizability of I$^-$ ions leads to stronger vdW interactions that effectively lock the lattice volume, maintaining it nearly constant until a significant substitution of iodine by bromine occurs. Consequently, a non-linear trend in the lattice parameter evolution is expected and justifies the observed deviation from Vegard's law. These stronger interactions likely restrict atomic rearrangement in the ab-plane while favoring ribbon elongation, resulting in highly anisotropic, needle-like morphologies in iodine-rich compositions.

*Table 1. DFT values of lattice parameters for the four Bi-chalcohalide parent compounds with the representative of the vdW interaction energy.*

|        | a(Å)  | b(Å)   | c(Å)  | vdW interaction energy (meV/ Å$^3$) |
|--------|-------|--------|-------|-------------------------------------|
| BiSI   | 8.342 | 10.024 | 4.144 | 7.39                                |
| BiSBr  | 8.056 | 9.518  | 4.032 | 6.25                                |
| BiSeBr | 8.079 | 10.704 | 4.083 | 7.05                                |
| BiSeI  | 8.550 | 10.304 | 4.190 | 8.19                                |

Raman spectroscopy was additionally performed on the Bi-based parent compounds and four of their solid solutions. The corresponding spectra is presented in **Fig. S3**. The Raman spectra can be divided into two distinct regions: vibrational modes below 150 cm$^{-1}$ are attributed to Bi–X (halogen) bonds, while those above 150 cm$^{-1}$ correspond to Bi–Ch (chalcogen) bonds.[35] In both regions, the solid solution samples exhibit broader peaks, with increased full width at half maximum (FWHM), relative to the parent compounds, indicating higher lattice disorder. Moreover, the absence of sharp, well-resolved peaks and the continuous shift of a single set of modes, rather than the coexistence of two distinct sets, provides strong evidence for the formation of a true solid solution, rather than a physical mixture of separate crystalline phases. This interpretation is consistent with the XRD results.

2.2. Elemental composition analysis

Elemental characterization of all Bi-chalcohalides, Bi(S$_z$,Se$_{1-z}$)(I$_x$,Br$_{1-x}$), was performed at the micrometre scale using Ion Beam Analysis (IBA) techniques. A combination of Particle-Induced X-ray Emission (PIXE) and Elastic Backscattering Spectrometry (EBS) were employed in a scanning nuclear microprobe to quantify elemental concentrations and assess depth profiles. PIXE enables high-precision elemental composition analysis without the need for calibration, detecting elements with atomic numbers above Z = 11 at ppm resolution.[42,43] EBS provides non-destructive in-depth information, enabling the determination of both layer thickness and composition.

PIXE 2D scanning maps of 113 x 113 μm$^2$ size were first performed in each sample to determine the elemental distribution homogeneity. Then, point analyses were carried out in selected regions to obtain a more precise elemental composition. In each measurement we first fit the EBS spectrum with a depth profile. Once this composition profile is correctly fitted, we use it as an initialization parameter in the PIXE spectrum fitting to accurately extract the bulk elemental concentration of the region selected.

**Fig. 3a-c** presents the EBS-fitted and PIXE spectra for BiS$_{0.3}$Se$_{0.7}$Br sample, and the elemental distribution in the 2D maps. The IBA measurements for all the studied compounds are presented in the **SI**. The 2D maps show the micrometre-scale homogeneity of the films, with no exposed substrate. Interestingly, in Se-containing samples, inhomogeneous regions are observed where the thickness of the active layer increases (the silicon signal from the Soda-Lime-Glass (SLG) decreases) and appear to be richer in bismuth and selenium. Therefore, these abnormal regions may originate from amorphous precipitates from bismuth-selenium complexes within the solution. The 2D-PIXE map, **Fig. 3c**, also reveals a uniform distribution of both sulphur and bromine in the film, with no distinguishable domains. This supports even more the formation of a true

solid solution, rather than separate crystalline domains as determined in the structural and vibrational characterizations (XRD and Raman).

Combining EBS and PIXE spectra from different point analyses the average elemental composition of the films was performed. **Table 2** presents the obtained atomic percentage of each element in the chalcohalide films. Although PIXE also detects elements from the substrate, the interlayer element diffusion observed in EBS spectra is attributed to morphology and rugosity effects rather than actual elemental diffusion (discussion in **SI**). Only Bi, S, Se, I, and Br were considered as part of the chalcohalide layer.

The obtained composition of the solid solutions closely aligns with the molecular precursor ratios in the formulated solutions. This confirms that for $BiSI_xBr_{1-x}$ and $BiS_zSe_{1-z}Br$, the intended x and z values in the precursor ink are effectively transferred to the final film composition without significant losses of material during the thermal treatment. These results highlight the strength of the methodology employed in this work, as it enables precise control over the film composition.

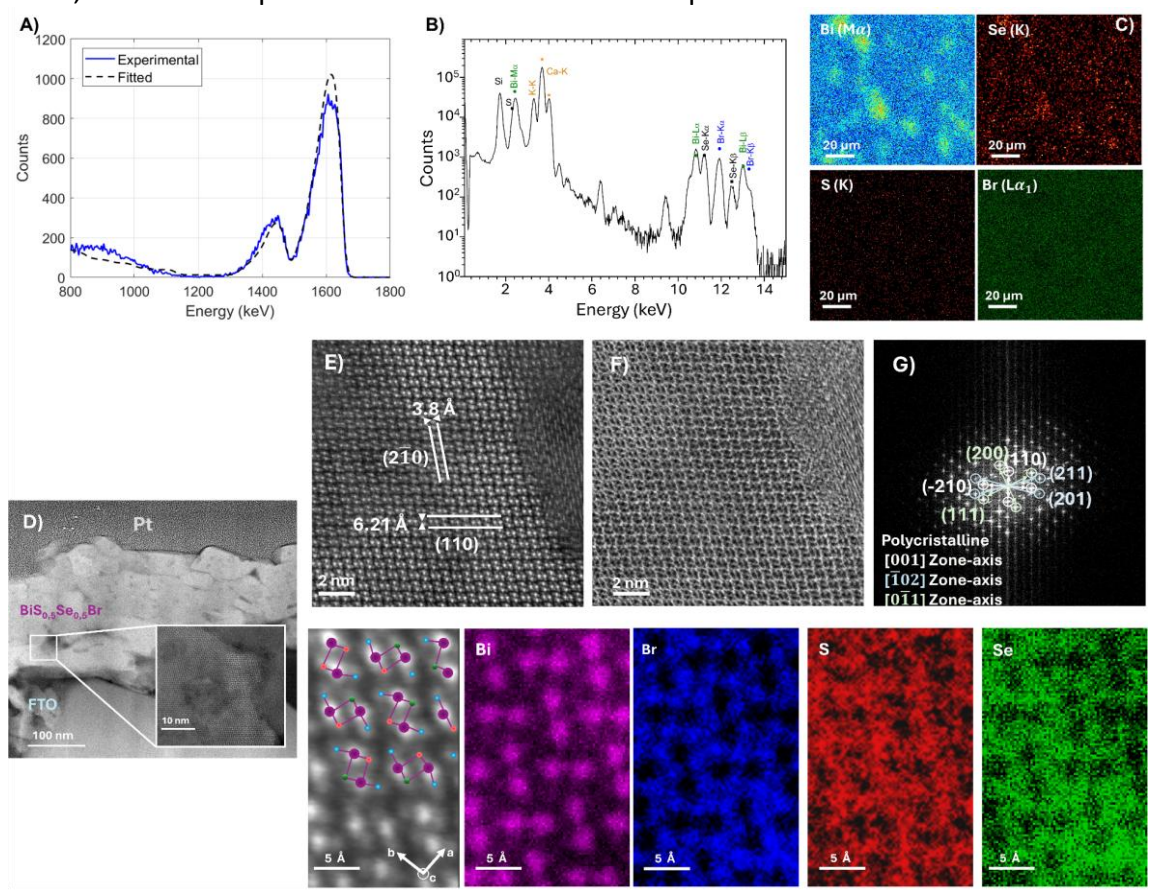

**Fig. 3. (a)** EBS (experimental and fitted), **(b)** PIXE spectra and **(c)** elemental maps (113 x 113 µm$^2$) for $BiS_{0.3}Se_{0.7}Br$ solid solution film on glass. **(d)** STEM cross-section of the $BiS_{0.5}Se_{0.5}Br$ on FTO with high-resolution image insert. **(e)** HAADF-STEM image showing ribbon-like atomic arrangement of heavy atoms (brightest spots being Bi atoms). **(f)** ABF-STEM image revealing the spatial distribution of lighter elements (S, Se, Br) around Bi atoms. **(g)** FFT pattern for the selected region. **Bottom:** HAADF-STEM image of with corresponding EELS elemental maps for Bi-M, Br-L, S-K and Se-L lines.

Hight-Resolution Transmission Electron Microscopy (HRTEM) was conducted on a Bi-chalcohalide solid solution to investigate its crystallinity and composition at the atomic scale. $BiS_{0.5}Se_{0.5}Br$ was selected for analysis due to the observation of certain inhomogeneities in the chalcogen solid solutions during IBA measurements. Consequently, this sample was examined to confirm the absence of compositional segregation. HRTEM analysis of the lamella revealed a polycrystalline structure, composed of chalcohalide Q-1D structures stacking both transversely and longitudinally (**Fig. 3d**). Transversal sections displayed single-crystalline domains with lateral dimensions corresponding to the Q-1D structure diameter, approximately 50 nm. In contrast, longitudinal sections, along the column axis, exhibited multiple crystalline domains stacked vertically. These observations suggest that each Q-1D structures consists of vertically stacked crystalline domains, with each domain spanning the full width of the column. This result provides insight into the film growth mechanism. Since the deposition involves multiple spin-coating steps to achieve the desired thickness, each layer can introduce a crystalline extension on top of the initial Q-1D structure, potentially with a different crystallographic orientation. As a result, a quasi-epitaxial growth of the columns along the vertical direction with stacking fault formation is proposed between spin-coating steps.

*Table 2. Elemental concentration of Bi-chalcohalide solid solutions obtained from PIXE spectra.*

|  | Bi (at%) | S (at%) | Se (at%) | I (at%) | Br (at%) |
|---|---|---|---|---|---|
| BiSI | 34.3 ± 0.11 | 33.9 ± 0.12 | - | 31.7 ± 0.01 | - |
| BiSI$_{0.7}$Br$_{0.3}$ | 35.6 ± 0.18 | 30.8 ± 0.68 | - | 23.4 ± 0.90 | 9.9 ± 0.37 |
| BiSI$_{0.3}$Br$_{0.7}$ | 32.9 ± 0.93 | 31.3 ± 1.54 | - | 9.8 ± 0.35 | 25.9 ± 1.63 |
| BiSBr | 32.9 ± 1.73 | 33.9 ± 1.82 | - | - | 33.0 ± 3.55 |
| BiS$_{0.7}$Se$_{0.3}$Br | 36.3 ± 0.76 | 23.2 ± 0.76 | 10.1 ± 0.47 | - | 30.3 ± 0.50 |
| BiS$_{0.3}$Se$_{0.7}$Br | 34.8 ± 0.82 | 10.6 ± 1.48 | 25.1 ± 1.62 | - | 29.4 ± 0.96 |
| BiSeBr | 40.5 ± 0.01 | - | 31.9 ± 0.01 | - | 27.5 ± 0.01 |
| BiSeI | 36.2 ± 0.47 | - | 26.1 ± 0.22 | 37.6 ± 0.53 | - |

**Fig. 3e-g** show magnified STEM maps of the region highlighted in **Fig. 3d**, along with the corresponding Fast Fourier Transform (FFT) pattern. From the FFT, interplanar distances of the crystalline planes were extracted and assigned to specific Miller indices. Strong agreement exists between the refined lattice parameters and the interplanar distances measured from the FFT. For example, the ($\bar{2}$10) plane, which has a calculated interplanar spacing of 3.76 Å based on the fitted diffraction data, appears with a spacing of 3.80 Å in the FFT pattern. Moreover, the FFT reveals signals from three different zone-axes: The (200) and (110) planes belonging to [0$\bar{1}$1] zone-axis, the ($\bar{2}$10) and (110) corresponding to [001] zone-axis and the (201) and (211) planes associated to [$\bar{1}$02] zone-axis. This indicates the presence of three crystalline domains within the selected region. In the HAADF-STEM image (**Fig. 3-Bottom**), the central crystalline domain exhibits an arrangement of atom pairs alternating in vertical and horizontal orientations, characteristic of the [001] zone-axis, viewed along the *c*-axis. These atom pairs represent the covalently bonded ribbons of the vdW structure, as projected onto the *ab* plane.

Electron Energy Loss Spectroscopy (EELS) measurements conducted on the crystalline domain oriented along the [001] zone axis (**Fig. 3-Bottom**) allowed direct visualization of the atomic distribution, where bismuth atoms form alternating vertical and horizontal pairs surrounded by lighter elements. Elemental maps for Bi, Br, S, and Se were obtained and show excellent agreement with the theoretical structure of the material (overlaid on the STEM image for comparison). Quantitative analysis of the EELS spectra was also performed to determine the elemental composition (**Fig. S4**). Three distinct regions, each measuring 15 Å x 30 Å, were analysed. The resulting average composition was 32 ± 5 at% Bi, 32 ± 3 at% Br, 17.1 ± 1.5 at% S, 16.7± 1.9 at% Se, which is in good agreement with the expected stoichiometry from the precursor solution, and are consistent with those obtained from IBA measurements. This confirms the formation of an actual solid solution, with stoichiometric composition with no segregation or formation of nanodomains.

    2.3.    Morphology control

Beyond precise control over composition and phase, the synthesis approach presented in this study enables systematic tuning of the morphology of chalcohalide compounds. As the optimal orientation and packing density of the material can vary depending on the target application, this morphological flexibility significantly enhances the potential for incorporating chalcohalides into diverse fields. For photovoltaic devices, compact and continuous thin films are essential to ensure proper top contact deposition and efficient extraction of photogenerated carriers (**Fig. 4b**). For sensing and catalysis, a high aspect ratio is preferred to maximize surface exposure and enhance interaction with the surrounding environment—making long, vertically aligned Q-1D structures more suitable (**Fig. 4b**). For thermoelectric applications, a horizontal alignment of the rods is desirable to optimize in-plane transport (**Fig. 4b**). Four parameters have been identified to play a critical role in this morphology tunning: substrate' skewness and kurtosis, $Bi(NO_3)_3·5H_2O$ : $BiX_3$ ratio in the precursor solution, spin-coating steps with different solution concentrations, and the solid-solution formulation (**Fig. 4a**).

The choice of substrate proves to be the most critical parameter for controlling the orientation of the chalcohalide films. Hydrophilic substrates are preferred, as they promote good wettability of the precursor solution, ensuring uniform coverage and film formation. However, if the contact angle is too low, the solution may spread too thinly during spin-coating, resulting in a very thin layer that leads to the formation of sparse and undersized chalcohalide rods after annealing. **Fig. S5** presents the measured contact angles and surface tensions for various substrates. Among them, ZnOS, CdS, $TiO_2$, FTO, and np-ZnO exhibit optimal surface tension that support desirable film morphology. In contrast, ZnSnO was found to be highly hydrophobic. Nevertheless, wettability alone does not fully account for the observed differences in morphology. As shown in **Fig. S6 and S7**, films deposited under identical conditions on $TiO_2$ and ZnOS substrates exhibit markedly different morphologies, despite having similar surface tensions. On CdS and ZnOS, the chalcohalide rods tend to grow more vertically, whereas on $TiO_2$ they tend to lie parallel to the substrate. Compared to $TiO_2$-based sample, the ZnOS substrate yields a more compact film, characterized by a merging of adjacent rods.

Atomic Force Microscopy (AFM) was performed on the substrates used for the growth of chalcohalide films (**Fig. S8**). Notable differences in surface morphology between substrates are observed. FTO, molybdenum, and $TiO_2$ exhibit relatively smooth, grainy surfaces, while CdS, ZnOS, and ZnSnO display grains with small protuberances. Nevertheless, the root mean square (RMS) roughness (**Table 3**) show no clear trend. For instance, FTO exhibits a relatively high RMS roughness of 8.99 nm, greater than that of CdS or ZnOS.

To better quantify these surface features, we explored additional statistical descriptors of the height distribution: skewness and kurtosis, which represent the asymmetry and sharpness of the height profile, respectively. A positive skewness suggests the presence of more surface peaks than valleys, while a kurtosis above 3 indicates sharp, narrow features that contribute to a non-uniform topography.[44]

As shown in **Table 3**, only CdS, ZnOS, ZnSnO, and np-ZnO exhibit both positive skewness and kurtosis values greater than 3, consistent with the visually observed surface protuberances. Notably, skewness is particularly pronounced for CdS, ZnOS, and ZnSnO, likely a consequence of their chemical bath deposition (CBD) synthesis method. These findings are further supported by the height distribution plots (**Fig. S9**), where these substrates show sharp, non-Gaussian profiles, in contrast to the more symmetric and smoother distributions seen for molybdenum or FTO.

Based on these results, it can be concluded that substrates with asymmetric height distributions—i.e., those exhibiting surface protuberances—promote the vertical alignment of chalcohalide rods by acting as preferential nucleation sites during the annealing process. Due to the sharpness of these surface features, localized evaporation of the precursor solution occurs more rapidly at these points, triggering early-stage crystallization of the chalcohalide rods (as illustrated in steps 2 and 3 of **Fig. 4d**).

Furthermore, a higher density of surface protuberances likely enhances the number of nucleation sites, which in turn leads to the formation of more compact thin films, as adjacent rods coalesce. This merging can be maximized through a two-step deposition strategy: first, a low-concentration precursor solution is used to grow small chalcohalide rods during annealing (step 4 in **Fig. 4d**), forming a seed layer; then, a second, higher-concentration solution is deposited, which promotes vertical and radial growth and merging into a dense compact film (step 5 in **Fig. 4d**). The seed layer formation is clearly observed in **Fig. S10** for the CdS and ZnOS substrates, where high crystalline grains are formed at the earlier growth stages.

Conversely, if a high aspect ratio and rod dispersion are desired—for instance, in sensing applications—a substrate with negative skewness and kurtosis below 3 should be chosen, as it offers fewer nucleation points and favours the isolated growth of rods.

*Table 3. Roughness, skewness and kurtosis values in the seven substrates studied obtained by AFM.*

|  | RMS (nm) | Skewness | Kurtosis |
|---|---|---|---|
| *Mo* | 7.38 | -0.197 | 3.49 |
| *FTO* | 8.99 | -0.18 | 3.70 |
| *np-ZnO* | 2.56 | 0.13 | 3.20 |
| *CdS* | 6.72 | 0.29 | 12.20 |
| *ZnOS* | 8.71 | 0.78 | 14.40 |
| *TiO$_2$* | 8.16 | -0.33 | 8.35 |
| *ZnSnO* | 12.26 | 0.19 | 7.01 |

Furthermore, noticeable morphological differences are observed across the various Bi-chalcohalides compositions. Compounds incorporating selenium tend to crystallize into thicker and more well-defined rods compared to their sulphur-based counterparts (**Fig. 4c**). In contrast, the halogen has a more pronounced influence on the shape rather than the size of the structures: iodine-based compounds typically form ribbon-like morphologies, whereas bromine-based compounds exhibit flake-like features. This confirms that the halogen significantly affects the crystal growth in the *ab*-plane, as shown previously by XRD analysis. As discussed previously, the lower polarizability of bromine likely weakens the vdW interactions between ribbons, promoting lateral rearrangement and facilitating a more isotropic growth. This results in the formation of more two-dimensional (2D) morphologies rather than Q-1D structures.

Beyond material composition, precursor solution formulation plays a crucial role in controlling the morphology, particularly influenced by the Bi(NO$_3$)$_3$·5H$_2$O : BiX$_3$ ratio (noted as X in **Table S1**). In the multistep spin-coating process, redissolution of earlier layers occurs in the absence of bismuth nitrate, mainly affecting smaller crystallized rods. As a result, only larger rods persist, serving as heterogeneous nucleation sites that preferentially grow during subsequent steps. This leads to the formation of isolated, large macroscopic rods sparsely distributed across the substrate (**Fig. S11**). Increasing the bismuth nitrate content and decreasing $BiX_3$ results in smaller rods and a denser, more compact film. This behavior can be attributed to two key factors: (1) the water introduced through bismuth nitrate pentahydrate decreases the solvation power of the solution, which helps preserve smaller rods from earlier layers and promotes a more uniform solute distribution during subsequent spin-coating steps; and (2) the reduced halogen content brings the precursor solution closer to stoichiometric proportions, thereby limiting abrupt halogen evaporation during crystallization at temperatures above 300 °C, which in turn minimizes material loss and enhances film integrity.

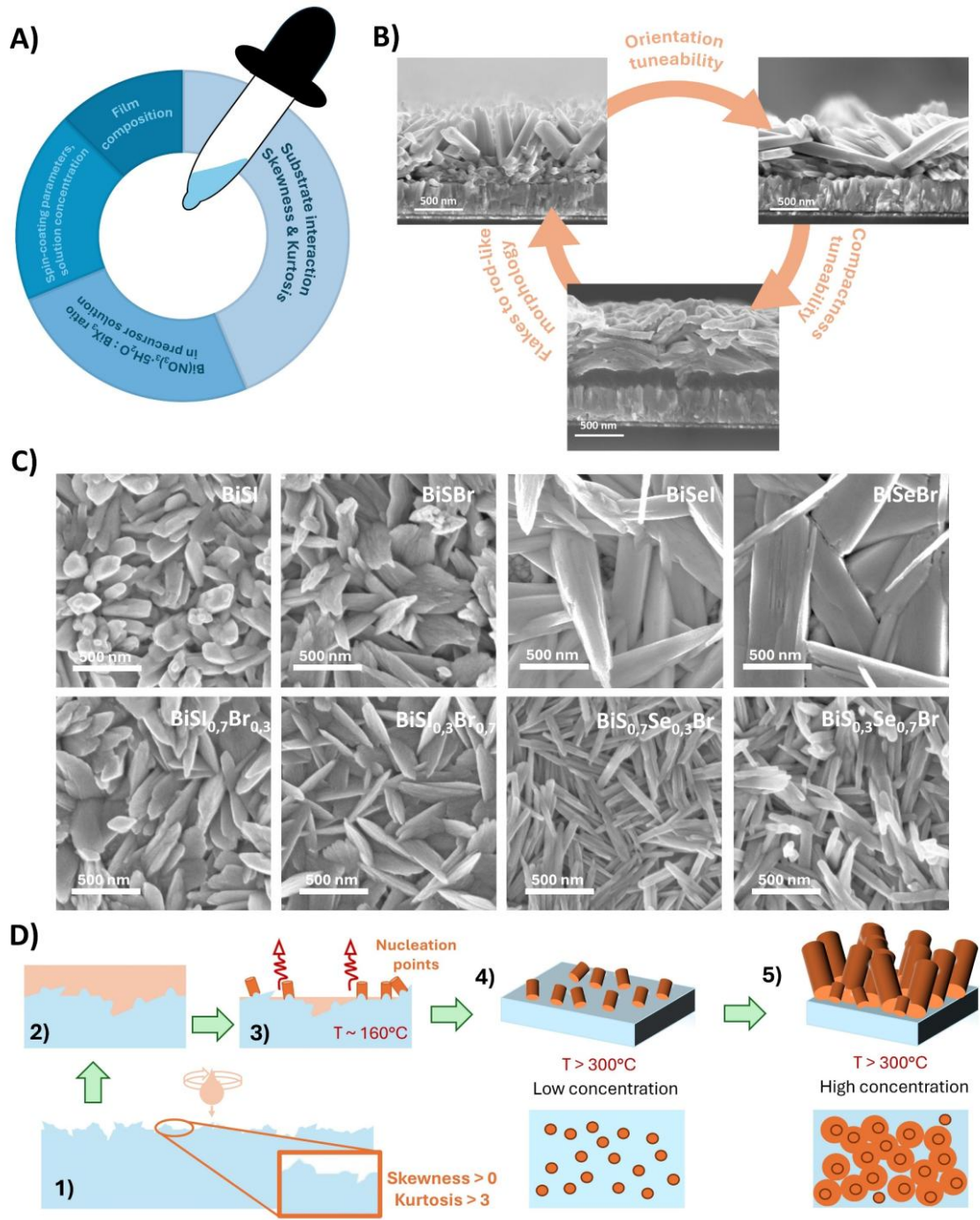

**Fig. 4. (a)** Diagram summarizing the key parameters controlling morphology, with different weights depending on the observed influence on crystal structure. **(b)** SEM cross-sectional images of BiSBr films grown under varying conditions reveal three morphologies: horizontally and vertically tilted ribbons, and compact film. **(c)** Top-view SEM images of Bi-chalcohalide solid-solutions on FTO/ZnOS substrates. **(d)** Solution-based synthesis method for compact films (substrates with skewness > 0 and kurtosis > 3). The process involves: **(1-2)** spin coating a low-concentration precursor solution, **(3)** mild annealing to nucleate grains at surface features, **(4)** high-temperature annealing to crystallize a seed layer, and **(5)** deposition of a higher-concentration solution, followed by crystallization to produce a dense film.

2.4. Optoelectrical properties

Photothermal Deflection Spectroscopy (PDS) was employed to determine the optical bandgap, assess the Urbach energy, and detect any possible sub-bandgap absorption features. Unlike UV-Vis spectroscopy, PDS allows the reliable detection of subtle spectral features such as Urbach tails, defect states, and sub-gap transitions.[45] **Fig. 5a-b** show the absorption coefficients derived from PDS measurements for Bi-chalcohalide solid solutions with varying halogen and chalcogen content, respectively. A systematic redshift in the absorption onset is observed with progressive anion substitution, indicating tunability of the bandgap. The optical band gaps and Urbach energies were extracted using a combined Urbach–Tauc fitting model incorporating both UV-vis and PDS datasets.[46,47] Through this approach the reliability of the extracted optical parameters is enhanced.

In the **SI**, we present the Tauc plots for all measured samples using both direct and indirect bandgap analysis methods. In each case, a linear trend is observed near the absorption onset, suggesting the coexistence of both direct and indirect band gaps in these chalcohalide materials, with the indirect band gap consistently lower in energy. These experimental results were compared with DFT calculations of the band structure showing good agreement. The DFT data confirm that these materials exhibit both direct and indirect transitions, with energy differences typically below 0.2 eV, and with the indirect bandgap always being the lowest.[48] Interestingly, while the indirect bandgap remains largely invariant with anion substitution, the direct bandgap shows greater sensitivity with a trend close to the one predicted by DFT (**Fig. 5c**). We observe a systematic tuning of the direct bandgap with halide and chalcogen substitution, in contrast to the indirect bandgap, which remains relatively unaffected. This tunability enables tailoring of the optical absorption edge, providing a pathway to optimize the material for specific spectral regions in light-harvesting devices.

**Fig. S13** shows the absorption coefficient fitted using the combined Urbach–Tauc model, which accounts for the coexistence of both direct and indirect band gaps. This dual-gap approach is crucial for accurately extracting the Urbach energy, as neglecting the indirect band gap can lead to a significant overestimation of the Urbach edge. The low Urbach energies observed for the parent compounds reflect their high structural and electronic quality. In particular, the iodine-containing compounds, BiSI and BiSeI, exhibit the lowest Urbach energies, 42 meV and 56 meV, respectively, indicating minimal disorder and high crystallinity. Furthermore, the absence of multistep absorption onsets supports the phase purity of these samples, with no detectable phase segregation. Upon anion substitution, the Urbach energy increases substantially, indicating a deterioration in lattice quality due to increased disorder in the solid solution. This effect is particularly pronounced in the chalcogen-substituted series: for example, Se incorporation into the BiSBr parent compound results in an order-of-magnitude increase in Urbach energy. This may be attributed to the preferential incorporation of chalcogen atoms into the covalent bonding direction, rather than the vdW gap, as suggested by Le Bail refinements. Such increased lattice distortion introduced by chalcogen substitution likely perturbs the electronic density of states near the band edges, which can negatively impact the optoelectronic quality of the semiconductor.

In the case of the halogen solid solutions $BiSI_{0.7}Br_{0.3}$ and $BiSI_{0.3}Br_{0.7}$, a sub-bandgap absorption feature is observed at approximately 1 eV, as highlighted in the logarithmic scale of **Fig. S13.** However, we exclude the possibility that this absorption corresponds to a true band edge transition, as it does not correlate with any linear region in the Tauc plots. To account for this feature, we included a Gaussian-shaped discrete energy state in the Urbach–Tauc fitting model of the absorption coefficient. In both compositions, this localized state was centered at 1.12 eV, suggesting the presence of a deep defect level that could potentially hinder charge transport and reduce device performance when the material is used in optoelectronic applications. Similarly, the work by Dolcet et al. on photoluminescence (PL) spectroscopy of BiSeI microcrystals revealed a pronounced deep emission band centred at $E_M$ = 1.108 eV in low-temperature PL measurements.[18] This emission closely matches the sub-bandgap absorption feature observed in our study, reinforcing the presence of a deep electronic state. López and collaborators[19] performed DFT calculations on the point defect landscape of BiSeI, identifying chalcogen vacancies as among the most detrimental intrinsic defects in Bi-based chalcohalides. These vacancies were predicted to introduce deep states with band-to-defect transition energies near 1 eV, in good agreement the sub-gap absorption observed experimentally. Interestingly, a similar sub-gap state was also detected in parent compounds synthesized using $Bi(NO_3)_3$-rich molecular inks, where it was accompanied by elevated Urbach energies (**Fig. S14**). Reducing the $Bi(NO_3)_3$ content in the precursor solution suppressed the discrete state and significantly lowered the Urbach energy. From XRD and SEM characterization one can observe that increasing the $Bi(NO_3)_3 \cdot 5H_2O$ content in the molecular ink reduces the crystallinity of the Q-1D chalcohalide structures, as reflected by broader XRD peaks and the development of more compact and dense morphologies. Since similar defect states are observed in solid-solution configurations where structural order is reduced, this suggests that crystalline distortion and partial amorphization promote the formation of deep defect states. Therefore, for the halogen solid solution series, implementing a post-deposition annealing step may help relax the structure and eliminate such defects, mirroring the improvements observed with optimized precursor formulations in parent compounds. Additionally, shifting the precursor composition toward chalcogen-rich conditions could further mitigate these defects, which have been theoretically attributed to sulfur or selenium vacancies.[19]

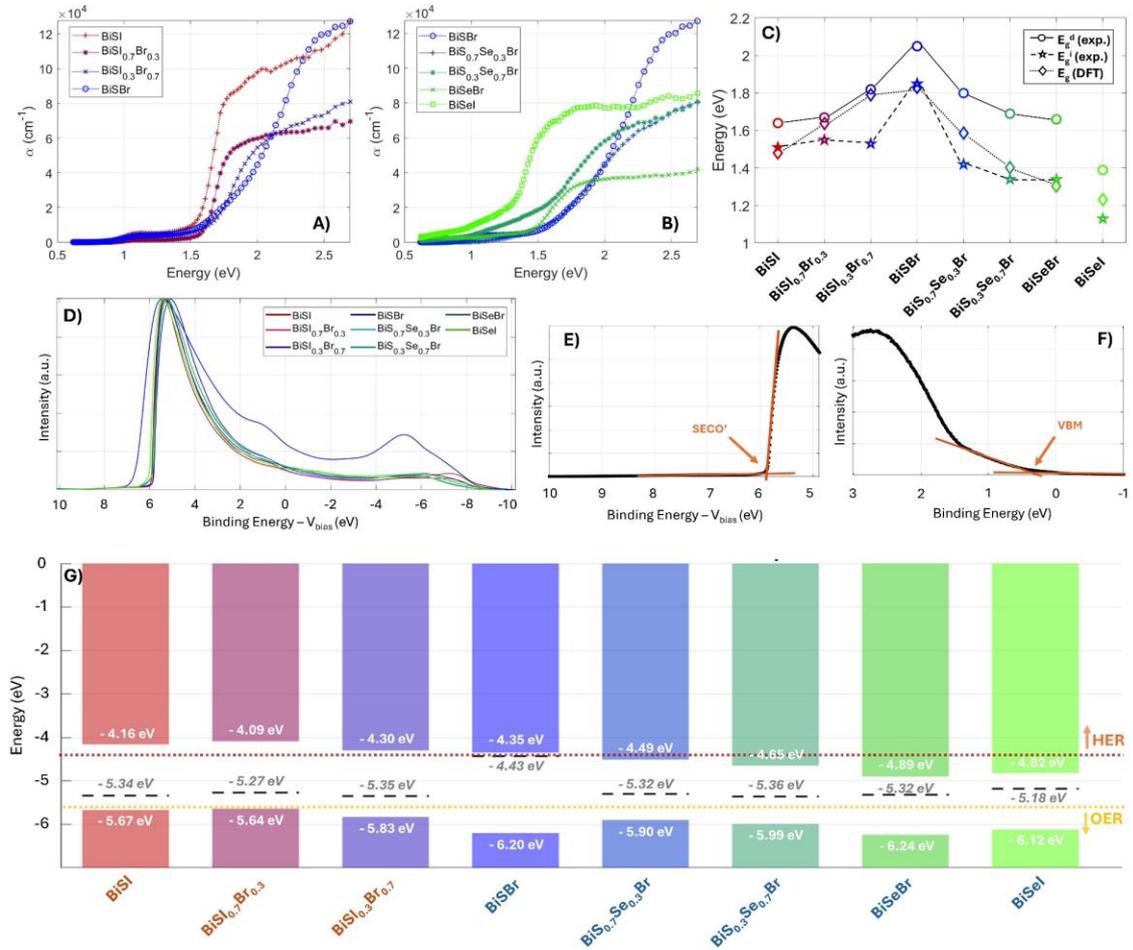

**Fig. 5.** Absorption coefficient spectra for **(a)** halogen and **(b)** chalcogen solid solutions obtained via PDS characterization. **(c)** Experimental and computational values for Bi-chalcohalides bandgaps, where "d" and "i" subindexes denote for direct and indirect bandgap. **(d)** UPS spectra for all Bi-chalcohalide compounds, measured with 10 eV bias voltage. **(e)** Magnified view of the bias UPS signal in the near vacuum level range with the graphical determination of SECO'. **(f)** Magnified view of the UPS signal in the near Fermi level range and the VBM graphical determination. **(g)** Energy band structure reconstruction for all Bi-chalcohalide compounds referenced to the vacuum level.

Ultraviolet Photoelectron Spectroscopy (UPS) was performed on all Bi-chalcohalide samples to analyze their electronic structure. Prior to measurement, the sample surfaces were gently sputtered for one minute to remove superficial oxidation without significantly altering the underlying semiconductor. The sputtering time was intentionally limited to minimize compositional changes learned from X-ray Photoelectron Spectroscopy (XPS) measurements that could arise from the differential sputtering yields of the elements involved, which could otherwise result in electronic structure modifications. **Fig. 5d** presents the UPS spectra of the samples, while **Fig. 5e** illustrates a representative example of the secondary electron cut-off (SECO') both measured under an applied bias voltage of 10 eV. **Fig. 5f** illustrate a representative example of the valence band maximum (VBM) in an unbiased sample.[49,50] Accordingly, the actual SECO is calculated as: SECO = SECO' + $V_{bias}$.

From these SECO and VBM values, the work function (WF) and ionization energy (IE) of the Bi-chalcohalide semiconductors were calculated using the following relationships: WF = hv – SECO and IE = hv – (SECO – VBM). By combining these UPS-derived values with the optical bandgaps obtained from PDS measurements, we constructed the electronic band diagrams for all studied Bi-chalcohalide compositions, as shown in **Fig. 5g**. The

sulphur–iodine compounds exhibit clear p-type conductivity, which progressively becomes more intrinsic as iodine is partially replaced by bromine. In the fully substituted BiSBr, the material shows a marked n-type behaviour, indicating a shift in Fermi level position across the solid solution series. Given its low Urbach energy, moderate bandgap (~1.6 eV), and p-type nature, BiSI emerges as the most promising Bi-chalcohalide candidate for optoelectronic devices. In contrast, the larger bandgap and strong n-type conductivity of BiSBr make it a suitable material for use as an electron transport layer (ETL) in various photovoltaic architectures.

Similar n-type conductivity is observed for the BiSeI and BiSeBr parent compounds. However, upon substituting sulphur with selenium in BiSBr, the material transitions toward a more intrinsic behaviour, accompanied by a downward shift of the conduction band minimum (CBM). This trend suggests that selenium orbitals dominate the conduction band density of states (DOS) and play a critical role in determining the electronic transport properties. This strong influence of selenium on the conduction DOS is consistent with DFT calculations reported by López et al.[48], reinforcing the excellent agreement between theoretical predictions and experimental observations for this class of materials. Moreover, these findings are further supported by XPS measurements, which show a shift in the Bi 4f peak (**Fig. S15-S16**), and XRD analysis (**Fig. S1**), which reveal greater lattice distortion along the covalently bonded directions in selenium-rich compositions. Together, these results highlight the pivotal role of the chalcogen anion— particularly selenium—in shaping the electronic structure and conduction properties of Bi-based chalcohalide semiconductors.

Moreover, analysis of the VBM and CBM positions reveals the potential suitability of these semiconductors for photoelectrochemical (PEC) water splitting.[35] Based on this band alignment analysis, BiSI, BiSI$_{0.7}$Br$_{0.3}$ and BiSI$_{0.3}$Br$_{0.7}$ emerge as promising candidates for photocathodes, while BiS$_{0.7}$Se$_{0.3}$Br, BiS$_{0.3}$Se$_{0.7}$Br, BiSeBr, and BiSeI are more suitable as photoanodes in PEC water splitting.[51,52] Interestingly, BiSBr possesses a band alignment that enables operation in both configurations, underscoring the tunable and versatile nature of chalcohalide semiconductors. These findings highlight the potential of this material family not only for photovoltaic applications but also for solar-driven fuel generation, broadening their relevance to emerging renewable energy technologies.

### 3. Concluding remarks

In this work, we present a novel low-temperature synthesis method based on molecular ink deposition, capable of producing a wide range of pnictogen chalcohalide materials with excellent morphological and compositional control. This methodology is notable for its simplicity, direct formation of ternary alloys, and the elimination of the need for pre-synthesized chalcogenide binary films. Using this approach, we explored the feasibility of forming solid solutions in Bi-based chalcohalides, motivated by their potential for defect engineering in next-generation high-efficiency devices. A thorough structural characterization confirmed the successful synthesis of phase-pure solid solutions and revealed a pronounced structural anisotropy intrinsic to their vdW layered nature. Additionally, we demonstrated precise compositional control, with a strong correlation between the molecular ink formulation and the final film composition, and no signs of phase segregation at either the micro- or nanoscale. Moreover, the ink-based deposition route is inherently scalable and compatible with large-area, solution-processing techniques, offering a pathway toward low-cost and industrially viable fabrication. Importantly, the molecular design concept underlying this approach is generalizable beyond Bi-based systems, enabling its adaptation to a broader class of chalcogenides and low-dimensional semiconductors.

Through comprehensive optoelectronic characterization, we were able to disentangle the individual roles of the chalcogen and halogen anions in tuning electrical conductivity and optical absorption, emphasizing the versatility and tunability of this material family. Notably, we identified the emergence of deep defect states upon introducing lattice disorder and proposed a passivation strategy based on post-synthesis thermal treatments for lattice relaxation. Beyond photovoltaics, we discussed the application potential of these materials in areas such as photoelectrocatalysis and thermoelectrics, enabled not only by their favourable optoelectronic properties, but also by the ability to precisely control their morphology. In this context, we identified the key parameters governing morphological evolution and demonstrated the growth of chalcohalides ranging from film-like layers to rod-shaped microcrystals.

Altogether, the fundamental insights and methodological advances presented here lay a solid foundation for the future integration of chalcohalides into high-efficiency optoelectronic and energy conversion technologies, including photovoltaics, photoelectrocatalysis, and thermoelectrics.

## 4. Materials and methods
### 4.1. Materials

Thiourea (> 99%), cadmium sulfate hydrate (98%), N,N-Dimethylformamide, and fluorine-doped tin oxide (FTO) substrates were purchased from Sigma-Aldrich. Bismuth (III) bromine powder (> 99%), bismuth (III) iodide powder (> 99.999%), bismuth (III) nitrate pentahydrate (> 99.999%), selenourea (98%), and Titanium(diisopropoxide)bis(2,4-pentanedionate) (75% in ethanol), ammonium hydroxide (28-30% in water), 2-propanol (> 99%) were purchased from ThermoScientific. Zinc sulfate heptahydrate (> 99.5%) was purchased from Chem-Lab. Liquinox critical cleaning liquid detergent was purchased from Alconex. Molybdenum substrates were purchased from Suzhou ShangYang Solar Technology. Zinc oxide nanoparticles (2.5% in mixture of alcohols) were purchased from Avantama.

### 4.2. Synthesis of Bi-chaclohalides

Film synthesis has been done through molecular ink deposition. A single molecular ink consisting of thiourea or selenourea ($SC(NH_2)_2$, $SeC(NH_2)_2$) as the chalcogen source, bismuth nitrate pentahydrated ($Bi(NO_3)_3·5H_2O$) as the Bi-salt, and bismuth iodide or bromide ($BiI_3$, $BiBr_3$) as the halogen salt is employed. These salts are sequentially dissolved in DMF in the order listed, in a way that the metal cations preferentially coordinate with the thione sulfur (or selenone) atoms rather than with the carbonyl oxygen of DMF. The total ratio of Bi/Ch in the solution is maintained at 1:1 to avoid the formation of the binary chalcogenide phases ($Bi_2(S,Se)_3$). However, the ratio of $Bi(NO_3)_3·5H_2O$ and $BiX_3$ can be adjust towards halogen-rich regions as the excess of halogen will be released during the temperature treatment (**Table S1**). Different concentrations have been studied for all the precursor species, ranging from 0.05 M to 1 M. After dissolution, 50 µL of the molecular ink is transferred onto the selected substrate using a micropipette. The substrates are pre-cleaned to ensure proper adhesion. The precursor is then spin-coated at 1000–2000 rpm for 30 seconds, after which the sample is transferred to a hotplate for soft annealing. A pre-annealing step is performed at 150 °C for 3 minutes to gently remove the solvent. Subsequently, the temperature is increased to the crystallization temperature of the material, 300-350 °C, and maintained for a minimum of 3 minutes. All the preparation is carried out in an inert atmosphere (using a glovebox system) with controlled oxygen and water levels (<1 ppm and <30 ppm, respectively).

### 4.3. Substrate preparation

Molybdenum and fluorine-doped tin oxide (FTO) coated glass substrates were initially cleaned manually using Liquinox and deionized water, followed by immersion in 2-propanol for 10 minutes in an ultrasonic bath. A final rinse with ultrapure deionized water was performed to ensure complete removal of contaminants. Various functional substrate layers were studied, most of which were deposited on top of FTO.

Titanium dioxide ($TiO_2$) layers were prepared via spray pyrolysis using a 5% ethanol solution of titanium(IV) diisopropoxide bis(2,4-pentanedionate), followed by thermal annealing at 500°C for 20 minutes. Zinc oxide (ZnO) layers were prepared by spin-coating ZnO nanoparticle slurry onto FTO, followed by annealing at 120°C for 10 minutes.

Cadmium sulphide (CdS) and zinc oxysulphide (ZnOS) layers were deposited via chemical bath deposition (CBD) on FTO substrates, using thiourea as the sulphur source and cadmium sulphate ($CdSO_4$) or zinc sulphate ($ZnSO_4$) as the metal precursor in an ammonia-based aqueous solution.

Finally, zinc tin oxide (ZnSnO) layers were grown on FTO by atomic layer deposition (ALD) using diethylzinc [$Zn(C_2H_5)_2$, DEZ], tetrakis(diethylamido)tin(IV) [$Sn(N(CH_3)_2)_4$, TDMASn], and deionized water as precursors. The deposition was performed at a substrate temperature of 150 °C under a continuous argon (Ar) carrier gas flow.

### 4.4. Material characterization

X-ray diffraction was performed in a PANalytical X'Pert PRO MPD powder diffractometer in Bragg-Brentano geometry, 240 millimetres of radius $\theta/2\theta$ goniometer, enabled to obtain the X-ray diffraction patterns, using nickel filtered Cu K$\alpha_{(1+2)}$ radiation, $\lambda_{average}$ = 1.5418 Å, with a work power of 45 kV – 40 mA. A divergence slit of 0.25 º, a mask defining a length of the beam over the sample in the axial direction of 12 millimetres and incident and diffracted beam Soller slits of 0.04 radians were used. An 1D PIXcel Detector with an active length of 3.347 º was employed. The measurements consisted of 2θ/ θ scans from 4,5 to 145º 2θ with a step size of 0.026º and a measuring time of 200 seconds per step. The patterns have been analysed with X'Pert HighScore software. Pattern matching refinement (via Le Bail method) has been performed using the FullProf suite and ensuring a weighted profile R-factor ($R_{wp}$) below 5%. The corresponding crystallographic information files (CIF) are available through the Crystallography Open Database (COD). The following COD entries were used as initial structures for the refinements: COD 1535800 (BiSI), COD 1535795 (BiSBr), COD 2010577 (BiSeI), COD 7711569 ($Bi_{12.89}Br_{2.85}S_{17.15}$), and COD 1000062 ($SnO_2$).

Raman spectroscopy measurements were carried out with a Renishaw inVia Qontor instrument with 532 nm excitation wavelength, and nominal 100 mW output power, coupled with 2400 lines/mm visible grating and Leica DM2700 microscope (with 100x objective) for confocal measurements. For data collection, between 0.5% to 5% of the nominal power was used. The optimal laser power was defined as the highest power at which no changes in the spectral parameters were detected. 20 accumulations with 3s of exposure time for each accumulation over 3 different areas of each sample were performed on each compound.

Ion beam analysis (IBA) measurements were carried out under vacuum conditions with the Oxford-type nuclear microprobe installed at the 2.5 MV Van de Graaff accelerator, in Lisbon (Portugal). For a detailed depth profiling analysis, the experiment was performed using two types of ion beams: protons at 1750 keV and alpha particles at 1800 keV. In both cases the beam current was ~300 pA. The use of alpha particles increases the elemental mass resolution and in-depth sensitivity when considering the EBS spectra. X-rays are detected using a 30 mm$^2$ Bruker SDD detector with a 145 eV resolution, positioned at 135° relative to the beam direction. When the proton beam is used, a 50 µm thick Mylar filter is placed in front of the detector. Backscattered particles were detected by means of a 200 mm$^2$ PIPS detector, positioned at 140° with the beam direction in Cornell geometry.

The OMDAQ software package was used for beam control, data acquisition, data processing and postdata treatment. PIXE and EBS maps were obtained by selecting the regions of interest (ROIs) in each spectrum using the list mode playback mode. PIXE spectra were analysed by the GUPIX code. For the quantification of elemental concentrations, Bi-M$\alpha_1$, S-K, Se-L$\alpha_1$, Br-K, I-L$\alpha_1$ shell emission lines were selected to their higher signal-to-noise ratio, ensuring more accurate calculations.

XPS measurements have been done with a monochromatic focused X-ray source (Aluminium K$\alpha$ line of 1486.6 eV) calibrated using the 3d$^{5/2}$ line of Ag with a full width at half maximum (FWHM) of 0.6 eV. The analysed area was a circle of 100 µm of diameter, sample placed at 45° with respect to the analyser axis, and the selected resolution for the spectra was 224 eV of Pass Energy and 0.8 eV/step for the general spectra, and 27 eV of Pass Energy and 0.1 eV/step for the high-resolution spectra of the selected elements. Measurements are referenced to the C$^{1s}$ signal, whose binding energy is equal to 284.8 eV in adventitious carbon (from atmospheric contamination).

On the other hand, UPS measurements have been done using a Helium source (He I line of 21.22 eV) calibrated using an Ag sample in which the work function (WF) calculated was of 4.27 eV. The analysed area was a spot of about 1.5 mm of diameter, sample placed at 90° with respect to the analyser axis, and the selected resolution for the spectra was 1.3 eV of Pass Energy and 0.01 eV/step. Both XPS and UPS measurements were made in an ultra-high vacuum (UHV) chamber at a pressure between 5x10$^{-10}$ and 5x10$^{-9}$ Torr. Moreover, to measure beyond the surface avoiding surface contamination sputtering with monoatomic Ar$^+$ ion gun (at 4 keV and 6mmx6mm raster) was done. The analysis and fitting of the spectra were carried out with Multipak V. 9.9.2 program.

The setup for Photothermal Deflection Spectroscopy (PDS) consists of a 100 W tungsten halogen lamp, PTI 01-0002 monochromator, and Thorlabs MC1000 optical chopper (4 Hz light modulation frequency). A Signal Recovery 7265 lock-in amplifier is connected to Hamamatsu C10442-02 PSD position sensitive detector to measure the deflection of the MC6320C laser probe beam (10 mW). Samples are located in a quartz cell filled with Fluorinert TM FC-40 liquid. UV-Vis data were also used to calibrate the PDS signal by normalizing the absorption coefficient in the high-absorption region. Optical analyses were performed by UV-Vis-NIR spectroscopy using a PerkinElmer Lambda 950 spectrophotometer operated with UV WinLab software. Measurements were carried out with a 150 mm integrating sphere, enabling global transmittance and reflectance acquisition over the 200–2500 nm wavelength range. The instrument is equipped with two detectors: a photomultiplier tube (PMT) for the UV-Vis region and a PbS detector for the NIR region.

Scanning electron microscopy was performed with a Zeiss Series Auriga field-emission microscope, with acceleration voltage of 5 kV and working distance ranging between 3 to 4 mm was employed to get top-view and cross-section images of the samples fabricated. No gold or carbon coatings were needed.

A VEECO Dimension 3100 equipment was used for the atomic force microscope measurements. The TAP150-G-10 probe with 10 nm tip radius, spring constat of 5 N/mm and resonant frequency of 150 kHz was acquired from Ted Pella, INC. Gwyddion software was used for the data treatment.

Water contact angle (WCA) measurements were conducted with a KRÜSS Drop Shape Analyzer (DSA25E). Precisely 1 µL of water was dropped onto the surface of the sample for every measurement. The acquired data were further examined utilizing the Sessile Drop module in the KRÜSS Advance program.

STEM-HAADF, STEM-EELS and STEM-EDXS data were acquired using a JEOL ARM 200cF (NEOARM), operated at 200 kV accelerating voltage. The microscope is equipped with a cold FEG electron gun and an aberration corrector in the condenser system. STEM-EELS data were acquired using a Gatan GIF Continuum K3 IS energy filter. STEM-EDXS data were acquired using the dual JEOL DrySD windowless detectors. The acquired data was processed through the software DigitalMicrograph from Gatan, INC. Plane identification in FFT were made using PDF reference 01-089-7110, with lattice parameters refined through Le Bail refinement for $BiS_{0.5}Se_{0.5}Br$: a = 4.035 Å; b = 10.05 Å; c = 8.12 Å.

### 4.5. DFT calculations

First-principles calculations based on density functional theory (DFT)[53] were carried out as implemented in the VASP code.[54] The projector augmented-wave method[55] was used to represent the ionic cores[56] and for each element the maximum possible number of valence electronic states was considered. Wave functions were represented in a plane-wave basis typically truncated at 850 eV. By using these parameters and a dense k-point grid for Brillouin zone integration of 10 × 5 × 4 centered at Γ, the resulting zero-temperature energies were converged to within 1 meV per formula unit. In the geometry relaxations, a tolerance of 0.005 eV·Å$^{-1}$ was imposed in the atomic forces.

To address the limitations of semi-local functionals,[57] we employed the range-separated hybrid functional HSEsol containing an exact Hartree-Fock exchange fraction of 25% [58,59], which is based on the Perdew–Burke–Ernzerhof exchange-correlation functional revised for solids.[60–62] Additionally, spin-orbit coupling (SOC) effects, which are particularly relevant for Bi-based compounds, were explicitly included[58,61,63] for the estimation of the optoelectronic properties, with a plane-wave basis truncated at 550 eV and a 5 × 3 × 2 Γ-centered mesh.

Solid solutions were modelled by means of the virtual crystal approximation (VCA)[64], which assumes the presence of virtual atoms on potentially disordered sites, interpolating between the electronic traits of the actual components. VCA supercells match the size of the primitive cell, making them computationally tractable while providing reliable results for isoelectronic atom substitutions.[48,65]

The strength of the long-range dispersion interaction in these materials was quantified using the vdW D3 correction scheme.[66] The energy of the vdW interaction was defined as the difference between the total ground energies calculated with and without the D3 correction. To facilitate the comparison across different chalcohalides, we performed single-shot energy calculations without structural relaxation, ensuring that the exact same atomic configuration (BiSI structure) was used for all chalcohalides. Thus, when normalizing any energy by units of volume we use the BiSI's unit cell volume.

## Acknowledgements

This research has received funding from the European Union H2020 Framework Program under the SENSATE project: Low-dimensional semiconductors for optically tunable solar harvesters (grant agreement Number 866018) and Renew-PV European COST action (CA21148). E.S. and J.P. acknowledge financial support of the Spanish Ministry of Science and Innovation from the CURIO-CITY project (PID2023-148976OB-C41), and ACT-FAST project (PCI2023-145971-2), from the CETP-Partnership Program 2022. This work was part of Maria de Maetzu Units of Excellence Programme CEX2023 - 001300-M, funded by MCIN/AEI/10.13039/501100011033.

D.R. acknowledges support by Generalitat de Catalunya under grant FI SDUR 2023 (ref. BDNS 699313) (DOGC Núm. 8929 – 02.06.2023). This research was funded by MCIN/AEI/10.13039/501100011033, grant number PID2022-138434OB-C53. Funding for the IBA measurements conducted in Lisbon, Portugal, was provided by *the European Union's Horizon Europe Research and Innovation programme under Grant Agreement No 101057511 (EURO-LABS). V.C. and L.C.A. acknowledge* Fundação para a Ciência e Tecnologia, Portugal.[67] J. W. T. and R. A. are grateful to the National Science Foundation for funding support through grants 1735282-NRT (SFEWS) and 10001536 (INFEWS). C.L. acknowledges support from the Spanish Ministry of Science, Innovation and Universities under a FPU grant. C.C. acknowledges support by MICIN/AEI/10.13039/501100011033 and ERDF/EU under grants TED2021-130265B-C22, TED2021-130265B-C21, PID2023-146623NB-I00, PID2023- 147469NB-C21 and RYC2018-024947-I and by the Generalitat de Catalunya under grants 2021SGR-00343, 2021SGR- 01519 and 2021SGR-01411. Computational support was provided by the Red Española de Supercomputación under grants FI-2025-2-0027, FI-2025-2-0028 and FI-2025-2-0006. Access to the JEOL ARM200F - NEOARM at CCiTUB was granted through the spanish Singular Scientific Technical Infrastructure (ICTS) ELECMI via a competitive call (project code ELC349-2023). E.S. acknowledges the ICREA Academia program.

## Research data for this article

Research data is available upon request. To request the data, contact the corresponding authors of the article.

## Conflicts of interest

The authors declare that they have no known competing financial interests or personal relationships that could have appeared to influence the work reported in this article.

# SUPPORTING INFORMATION

-------

## Molecular ink-based synthesis of Bi(S$_z$Se$_{1-z}$)(I$_x$Br$_{1-x}$) solid solutions as tuneable materials for sustainable energy applications

D. Rovira[1,2], I. Caño[1,2], C. López[2,3], A. Navarro-Güell[1,2], J.M. Asensi[4,5], L. Calvo-Barrio[6,7], L. Garcia-Carreras[6], X. Alcobe[6], L. Cerqueira[8], V. Corregidor[8], Y. Sanchez[9], A. Jimenez-Arguijo[1,2], O. El Khouja[1,2], S. Lanzalaco[2,12], J. W. Turnley[10,11], R. Agrawal[10], C. Cazorla[2,3], J. Puigdollers[1,2], E. Saucedo[1,2]

[1]Universitat Politècnica de Catalunya (UPC), Photovoltaic Lab – Micro and Nano Technologies Group (MNT), Electronic Engineering Department, EEBE, Av Eduard Maristany 10-14, Barcelona 08019, Spain
[2]Universitat Politècnica de Catalunya (UPC), Barcelona Centre for Multiscale Science & Engineering, Av Eduard Maristany 10-14, Barcelona 08019, Spain
[3]Universitat Politècnica de Catalunya (UPC), Group of Characterization of Materials (GCM), Physics Department, EEBE, Av Eduard Maristany 10-14, Barcelona 08019, Spain
[4]Universitat de Barcelona (UB), Applied Physics Department, C. Martí I Franquès 1, Barcelona 08028, Spain
[5]Institute of Nanoscience and Nanotechnology (IN2UB), Universitat de Barcelona, 08028 Barcelona, Spain.
[6]Universitat de Barcelona (UB), Scientific and Technological Centers (CCiTUB), C. Lluís Solé i Sabaris 1-3, Barcelona 08028, Spain
[7]Universitat de Barcelona (UB), Electronics and Biomedical Engineering Department, C. Martí I Franquès 1, Barcelona 08028, Spain
[8]Centro de Ciências E Tecnologias Nucleares (C2TN), Instituto Superior Técnico, Universidade de Lisboa, E.N. 10 Ao Km 139.7, 2695-066, Bobadela LRS, Portugal
[9]Catalonia Institute for Energy Research – IREC, Jardins de les Dones de Negre 1, Sant Adrià de Besòs 08930, Spain
[10]Davidson School of Chemical Engineering, Purdue University, West Lafayette, Indiana 47907, USA
[11]Department of Chemical Engineering and Materials Science, Michigan State University, East Lansing, Michigan 48824, USA
[12]Universitat Politècnica de Catalunya, Innovation in Materials and Molecular Engineering (IMEM), Department of Chemical Engineering, Av Eduard Maristany 10-14, Barcelona 08019, Spain

**Table S1.** Specification of precursor ratios for each Bi-chalcohalide compound. The values presented correspond to the ratio Bi(NO$_3$)$_3$·5H$_2$O : SC(NH$_2$)$_2$ : SeC(NH$_2$)$_2$ : BiI$_3$ : BiBr$_3$ . X denotes the proportion of Bi-precursor salt that corresponds to bismuth nitrate, X ∈ [0,1].

|  | *Bi(NO$_3$)$_3$·5H$_2$O* | *SC(NH$_2$)$_2$* | *SeC(NH$_2$)$_2$* | *BiI$_3$* | *BiBr$_3$* |
|---|---|---|---|---|---|
| *BiSI* | X | 1 | - | 1-X | - |
| *BiSI$_{0.7}$Br$_{0.3}$* | X | 1 | - | 0.7(1-X) | 0.3(1-X) |
| *BiSI$_{0.5}$Br$_{0.5}$* | X | 1 | - | 0.5(1-X) | 0.5(1-X) |
| *BiSI$_{0.3}$Br$_{0.7}$* | X | 1 | - | 0.3(1-X) | 0.7(1-X) |
| *BiSBr* | X | 1 | - | - | (1-X) |
| *BiS$_{0.7}$Se$_{0.3}$Br* | X | 0.7 | 0.3 | - | (1-X) |
| *BiS$_{0.5}$Se$_{0.5}$Br* | X | 0.5 | 0.5 | - | (1-X) |
| *BiS$_{0.3}$Se$_{0.7}$Br* | X | 0.3 | 0.7 | - | (1-X) |
| *BiSeBr* | X | - | 1 | - | (1-X) |
| *BiSeI* | X | - | 1 | (1-X) | - |

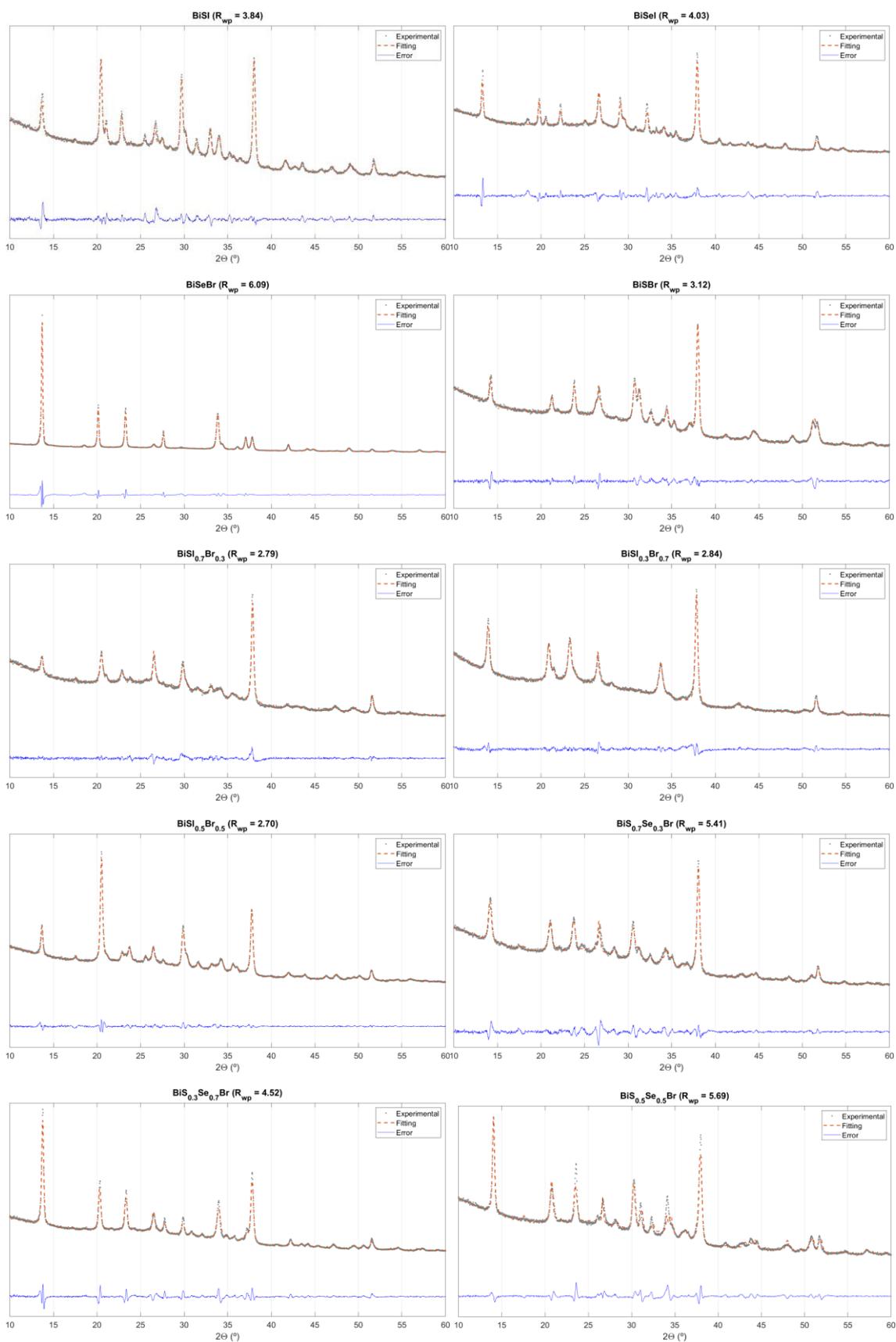

**Fig. S1.** Experimental X-ray diffraction patterns (grey) with corresponding Le Bail refined fits (orange) for all Bi-chalcohalide solid solutions. The difference between experimental and calculated profiles is shown in blue. The weighted profile R-factor ($R_{wp}$) for each refinement is indicated in the respective graph titles.

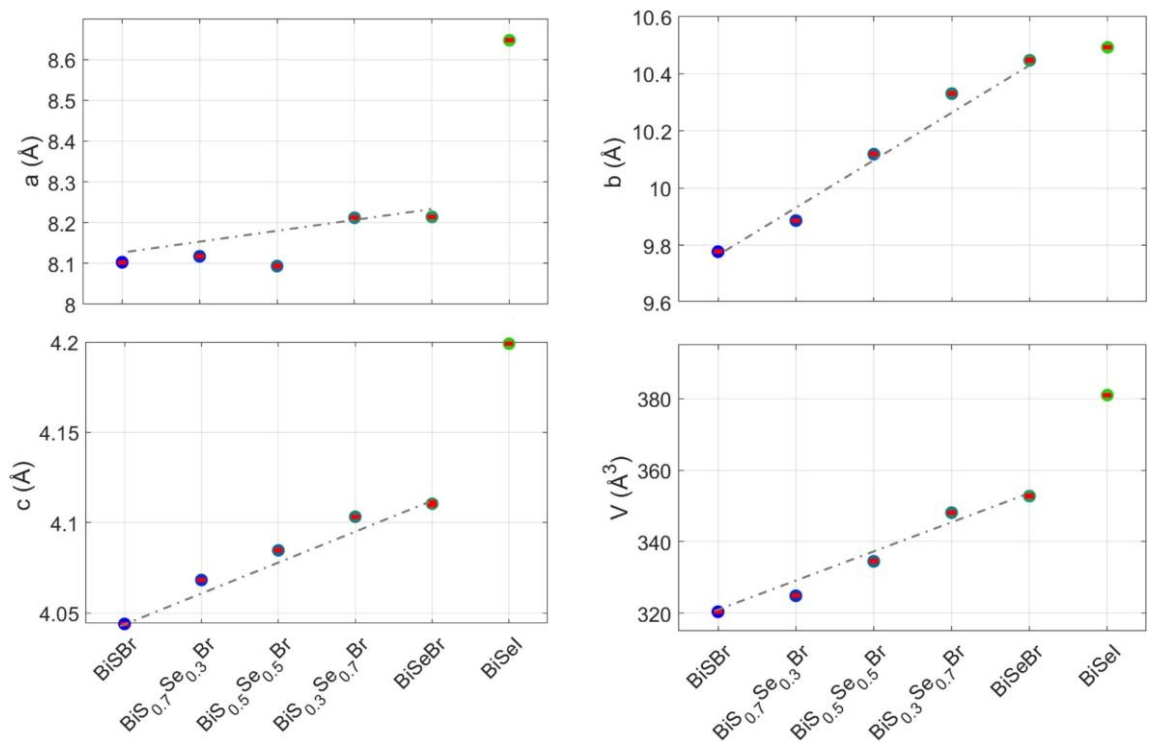

**Fig. S2.** Cell parameters and unit cell volume for the chalcogen solid solutions in Bi-chalcohalides. The dashed line indicates the theoretical evolution following Vegard's Rule. Error bars shown in red.

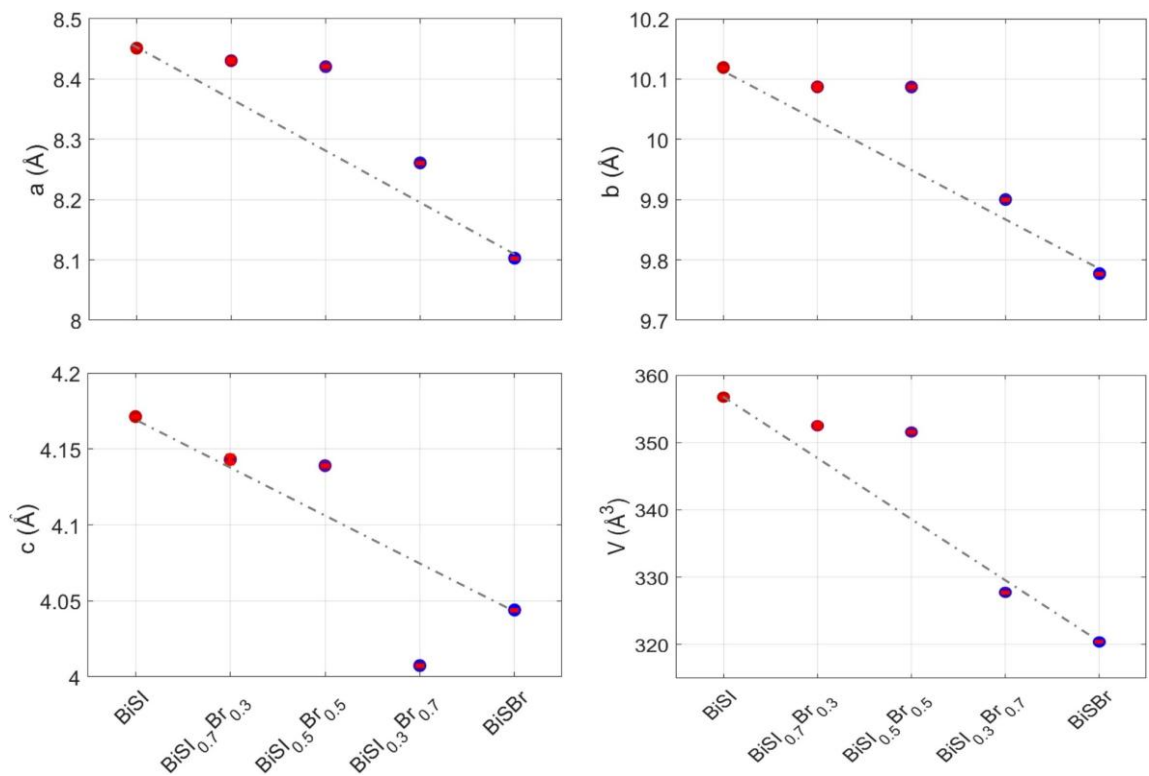

**Fig. S3.** Cell parameters and unit cell volume for the halogen solid solutions in Bi-chalcohalides. The dashed line indicates the theoretical evolution following Vegard's Rule. Error bars shown in red.

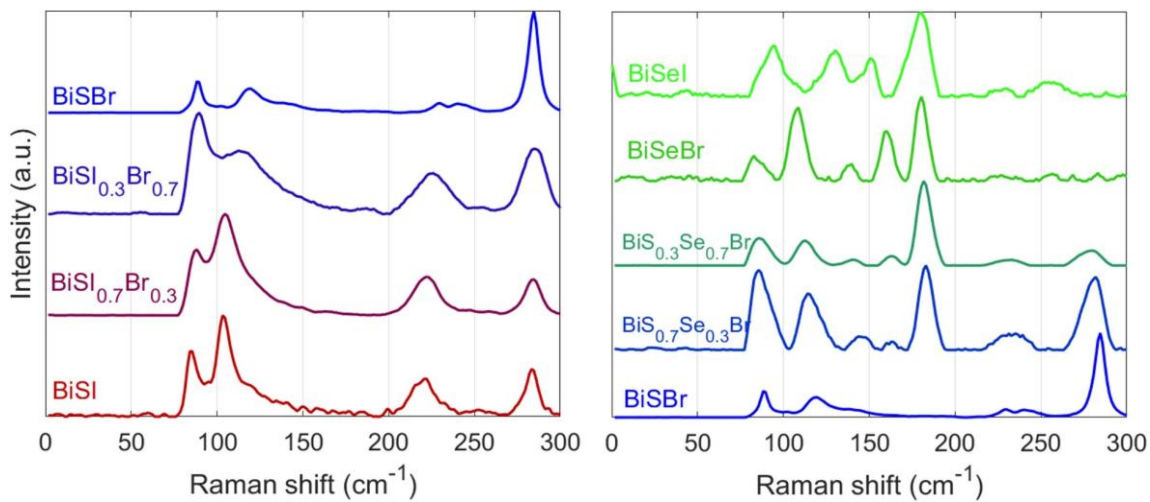

**Fig. S4.** Raman spectra for Bi-chalcohalide solid solutions excited with a 532 nm laser.

*IBA Analysis*

The EBS fitting was performed considering the columnar morphology of these compounds. Due to film deposition on substrates with weaker adhesion, such as SLG or FTO, the films exhibited lower compactness compared to those grown on other substrates. This aspect is further discussed in the **Morphology Control** section. As a result, instead of forming a fully compact film, the material exhibits a rod-like dispersion with empty spaces at the substrate interface (see **Fig. S19**). Consequently, the conventional approach of modelling the EBS spectrum as a simple layered stack (e.g., SLG/BiSI) was not applicable. Instead, the porosity of the active layer at the interface had to be considered. This porosity leads to reduced shadowing of the substrate's backscattered particles, causing the substrate signal to appear more pronounced in regions where the rods create more voids. Since the detected signal represents an average over areas with varying porosity, the EBS spectrum was fitted using a multilayer stack model of the form: SLG / $(SLG)_{0.95}(BiSI)_{0.05}$ /… / $(SLG)_{1-y}(BiSI)_y$ / … / $(SLG)_{0.05}(BiSI)_{0.95}$ / BiSI, with $y \in [0,1]$. When extended to an infinite number of layers, this approach effectively translates the interface porosity into a composition gradient in the EBS spectrum, resembling interlayer diffusion. Additionally, porosity leads to a reduction in the effective thickness of the layer as measured by EBS. Although cross-sectional scanning electron microscope (SEM) views (**Fig. S19**) indicates a film thickness of approximately 500 nm, the effective thickness interacting with backscattered particles in the EBS characterization is below 300 nm, as it depends on the total mass of material rather than the physical thickness alone.

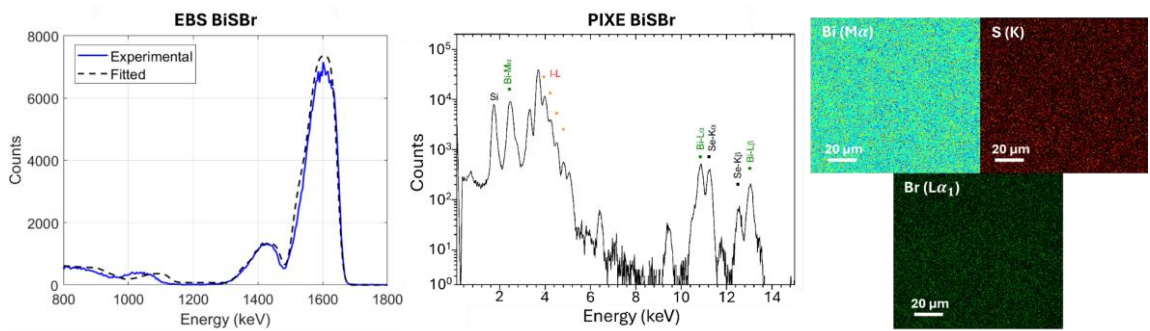

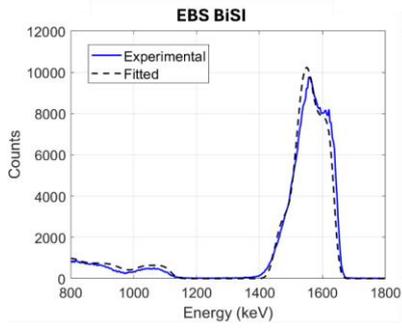 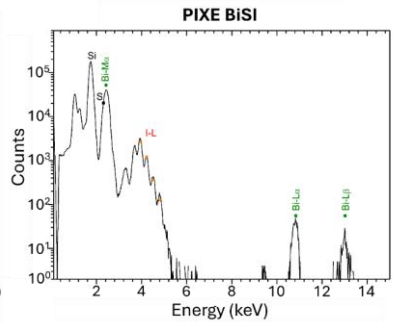 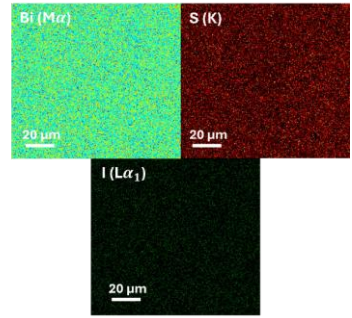
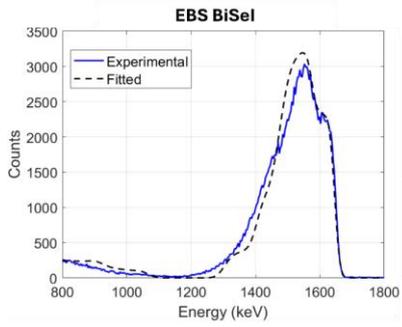 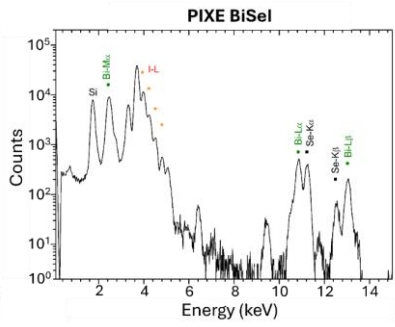 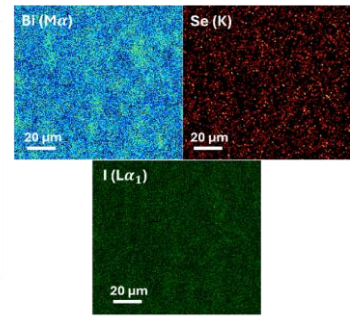
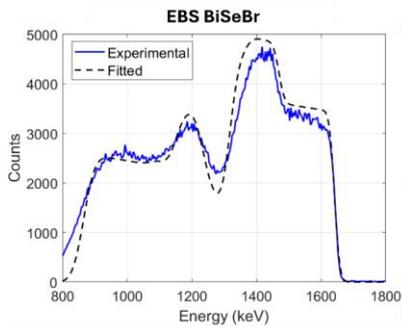 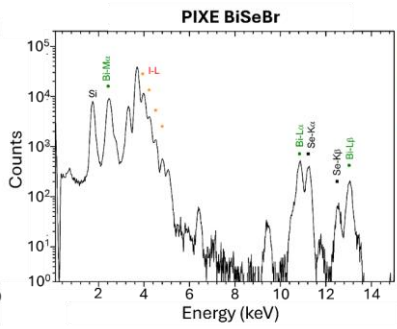 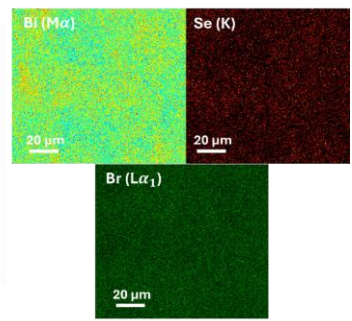
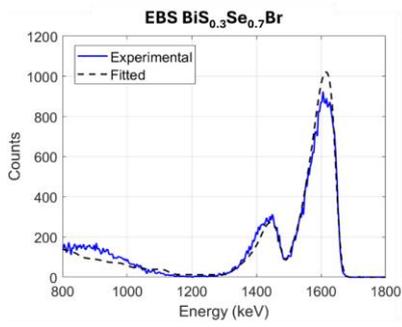 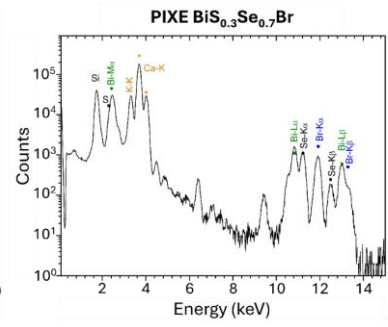 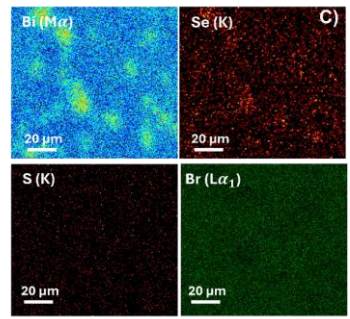
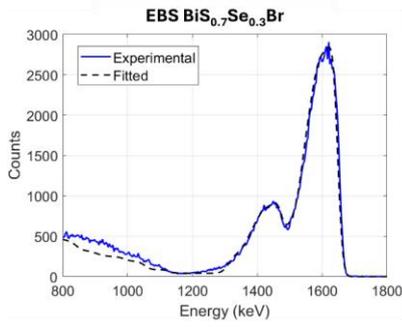 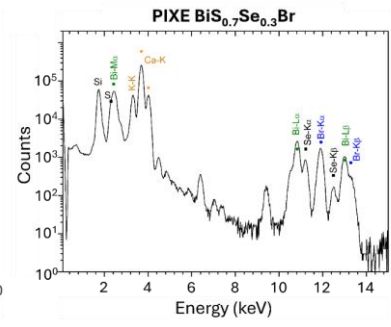 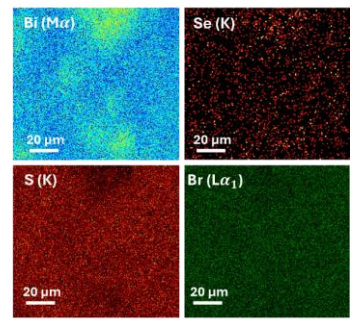

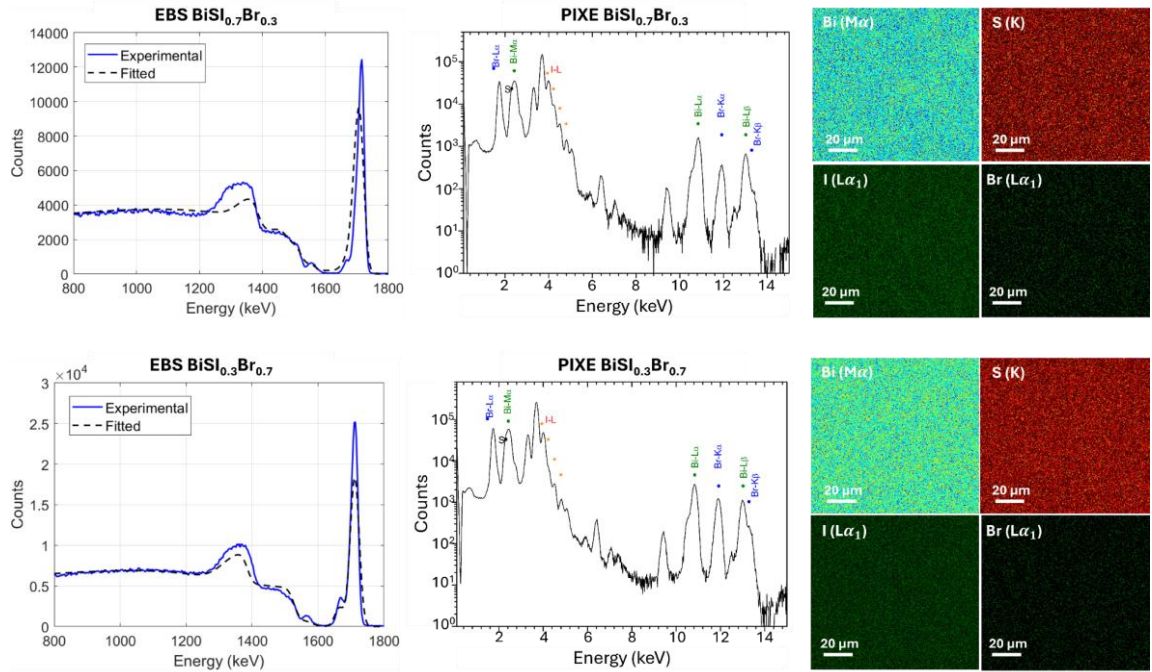

**Fig. S5.** From left to right: EBS spectra (experimental and fitted), PIXE spectra with main elemental peaks labeled, and corresponding elemental maps for Bi-chalcohalide films on glass.

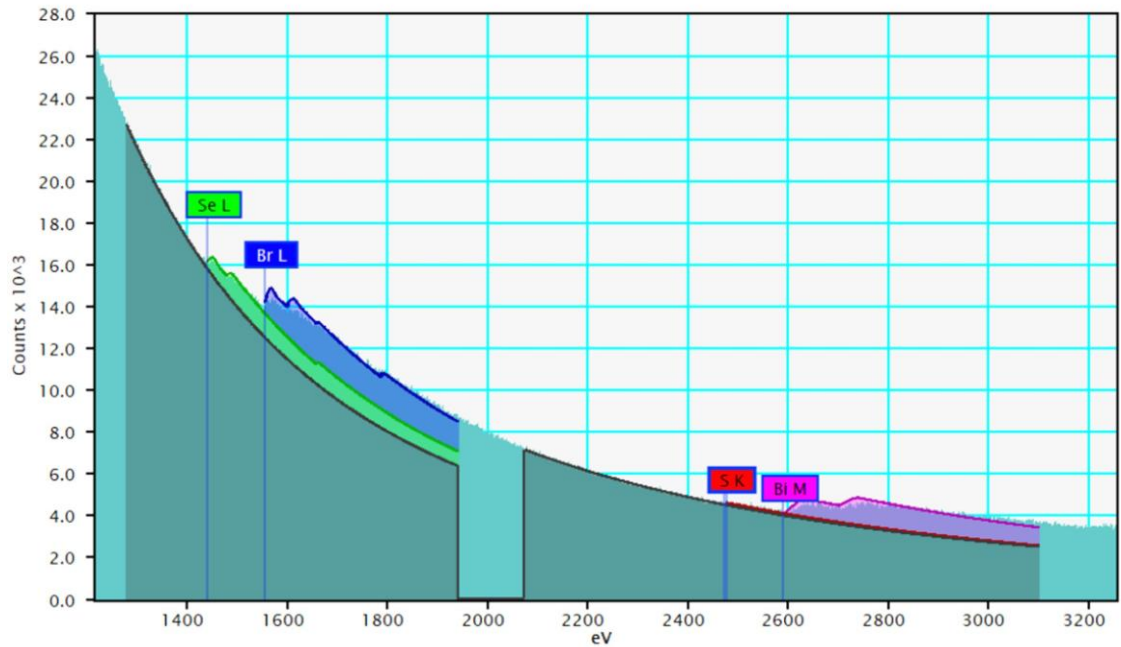

|      | Signal ( Counts )      | Comp. ( at.% ) | Comp. ( wt.% ) | X-section ( barns )   | X-section Model |
|------|------------------------|----------------|----------------|-----------------------|-----------------|
| Se L | 503.0e+03 ± 2.0e+03    | 16.7 ± 1.9     | 11.5 ± 1.3     | 3.3e+03 ± 0.3e+03     | Hartree-Slater  |
| Br L | 663.0e+03 ± 2.0e+03    | 32 ± 3         | 22 ± 2         | 2.2e+03 ± 0.2e+03     | Hartree-Slater  |
| S K  | 61.9e+03 ± 1.7e+03     | 17.1 ± 1.5     | 4.8 ± 0.4      | 393 ± 20              | Hartree-Slater  |
| Bi M | 537.8e+03 ± 1.5e+03    | 32 ± 5         | 61 ± 8         | 1700 ± 300            | Hartree-Slater  |

**Fig. S6**. EELS spectrum with elemental quantification for BiS$_{0.5}$Se$_{0.5}$Br.

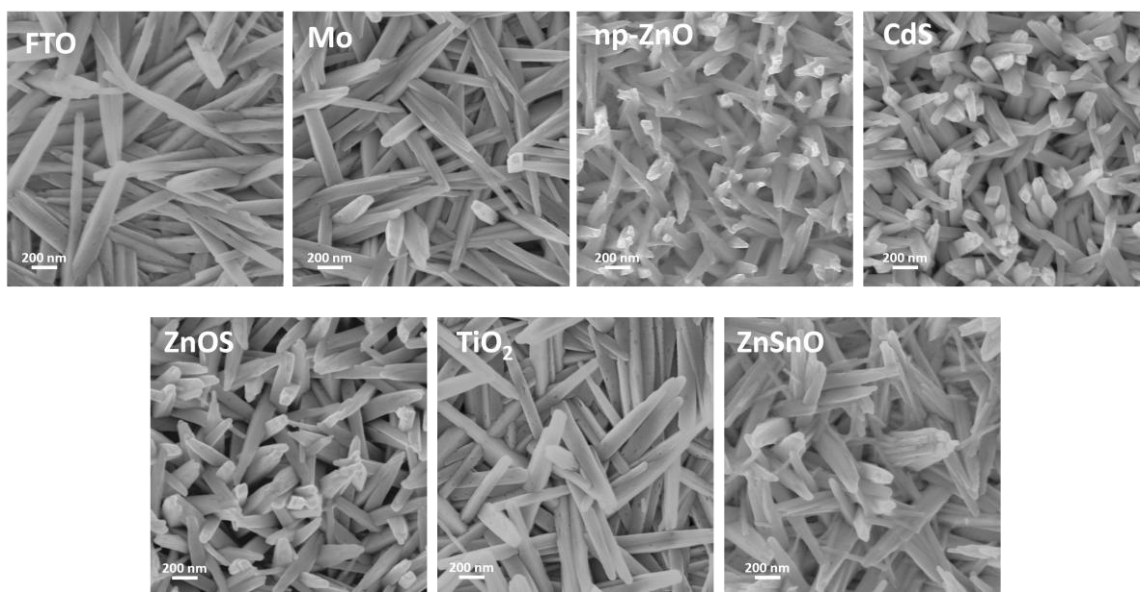

| Mo | np-ZnO | CdS | FTO | ZnOS | TiO$_2$ | ZnSnO |
|---|---|---|---|---|---|---|
| 18.0 ± 3.3° | 49.8 ± 2.1° | 50.2 ± 2.5° | 52.1 ± 2.2° | 69.8 ± 4.7° | 70.1 ± 3.3° | 111.1 ± 2.6° |
| 69.5 mN/m | 53.8 mN/m | 53.6 mN/m | 52.3 mN/m | 41.83 mN/m | 41.6 mN/m | 16.5 mN/m |

**Fig. S7.** Contact-angle measurements for seven different substrates grown over SLG.

**Fig. S8.** Top-view SEM images of BiSBr grown over different substrates after 3 spin-coating steps with a precursor solution concentration of 0.5 M.

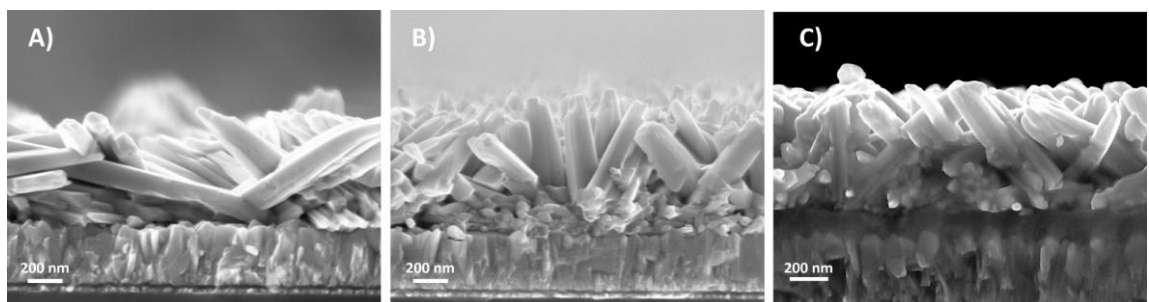

**Fig. S9**. Cross-section SEM images of BiSBr grown over **(a)** FTO/TiO$_2$, **(b)** FTO/CdS and **(c)** FTO/ZnOS with a Bi(NO$_3$)$_3$·5H$_2$O : BiX$_3$ ratio of 3:2.

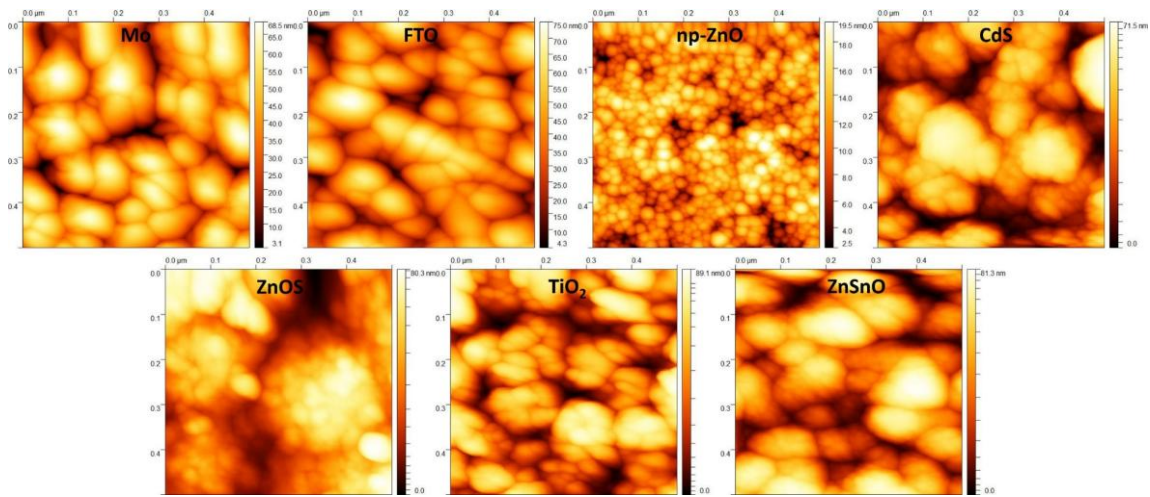

**Fig. S10.** AFM heigh images of different substrates used to grow Bi-chalcohalides.

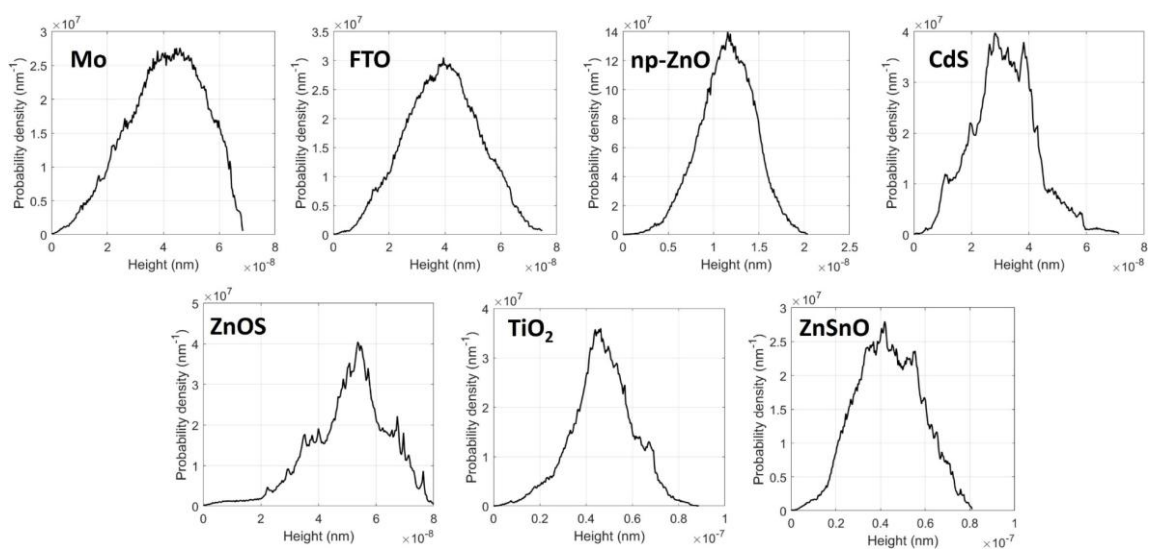

**Fig. S11.** Heigh distribution obtained from AFM measurements of different substrates' surface.

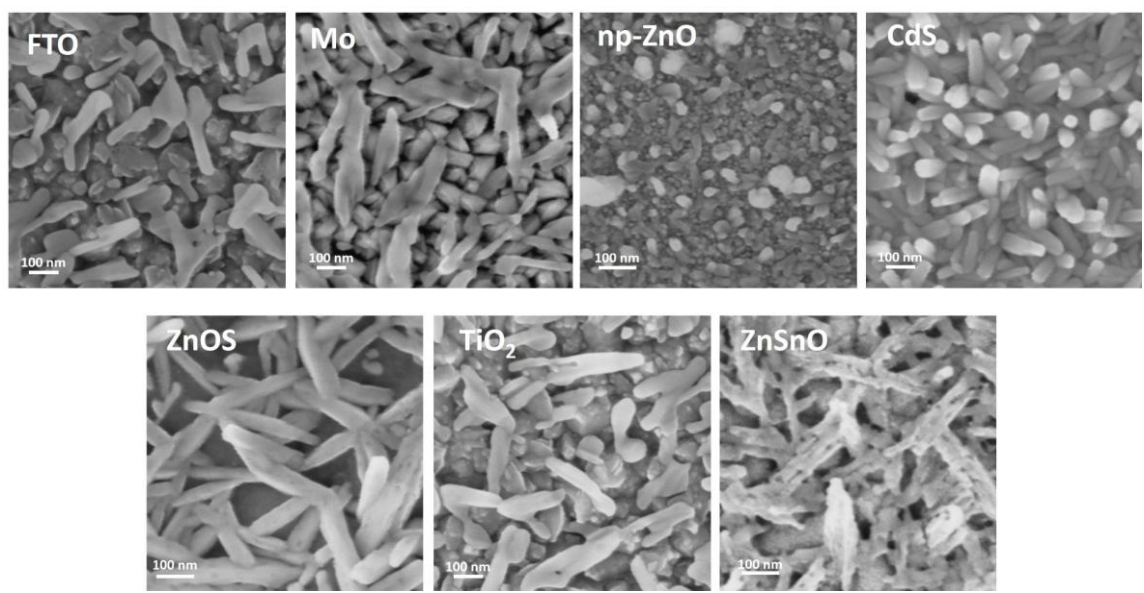

**Fig. S12.** Top-view SEM images of BiSBr grown over different substrates after 3 spin-coating steps using a precursor solution concentration of 0.05 M.

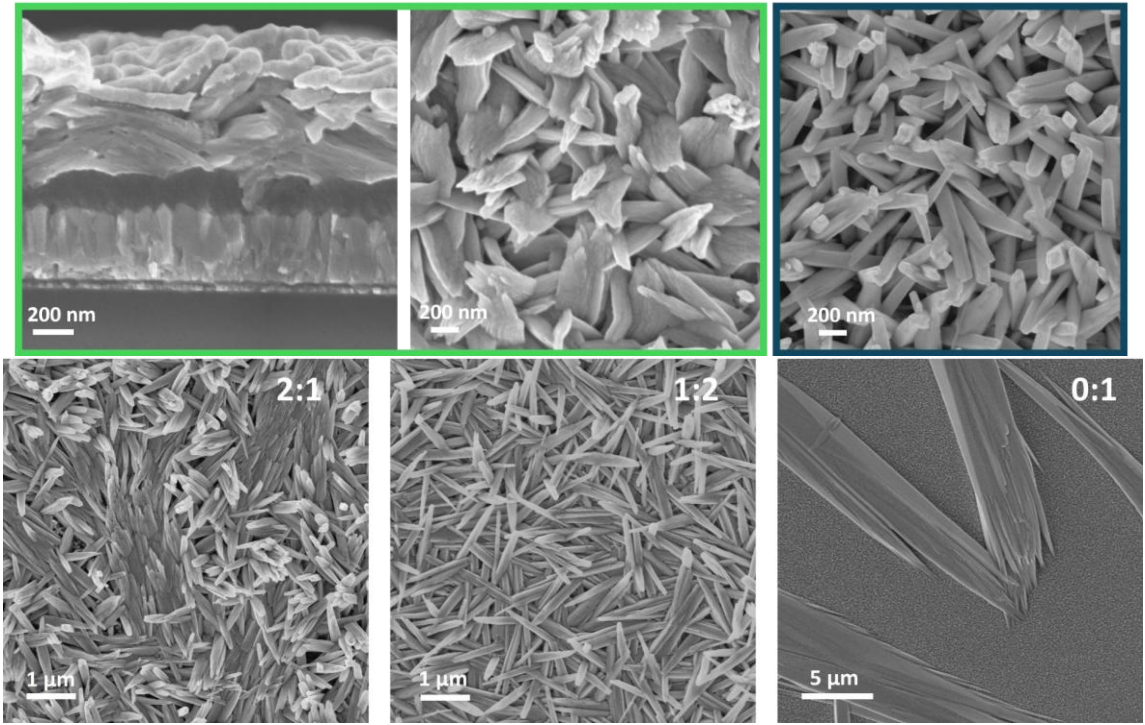

**Fig. S13. (Top)** Top-view and cross-sectional SEM images of BiSBr grown over ZnS with different Bi(NO$_3$)$_3$·5H$_2$O : BiX$_3$ ratios. **(Bottom)** Top-view SEM images of BiSBr grown over FTO with different Bi(NO$_3$)$_3$·5H$_2$O : BiX$_3$ ratios.

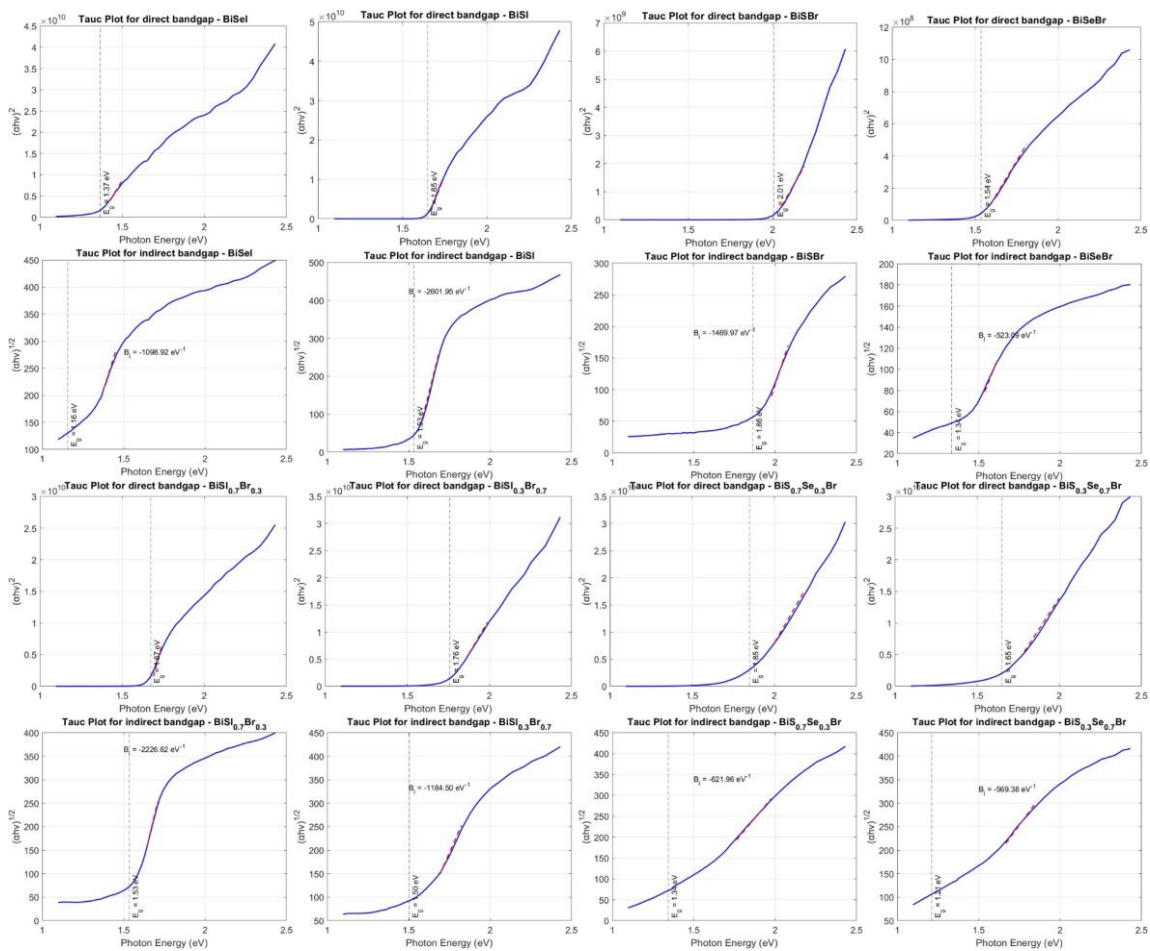

**Fig. S14.** Tauc plots for indirect and direct models from PDS absorption measurements of Bi-chalcohalide compounds.

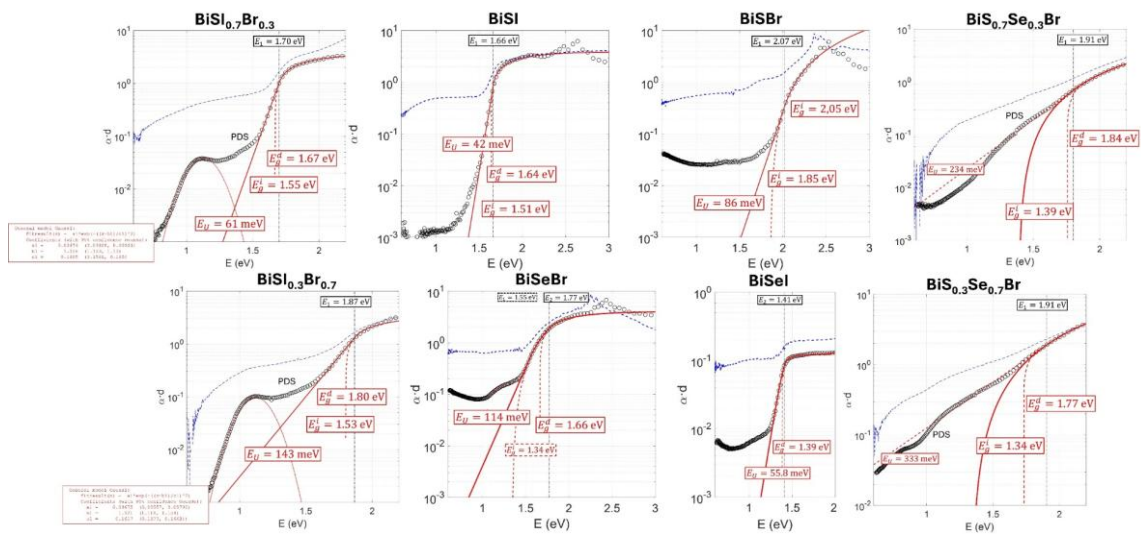

**Fig. S15.** PDS absorption fitted with Urbach-Tauc model. Experimental data in black, fitted model in red and optical absorption in blue. The values for the bandgaps and Urbach energy are displayed. For the halogen solid solutions we add the details of the gaussian fitted sub-gap discrete state.

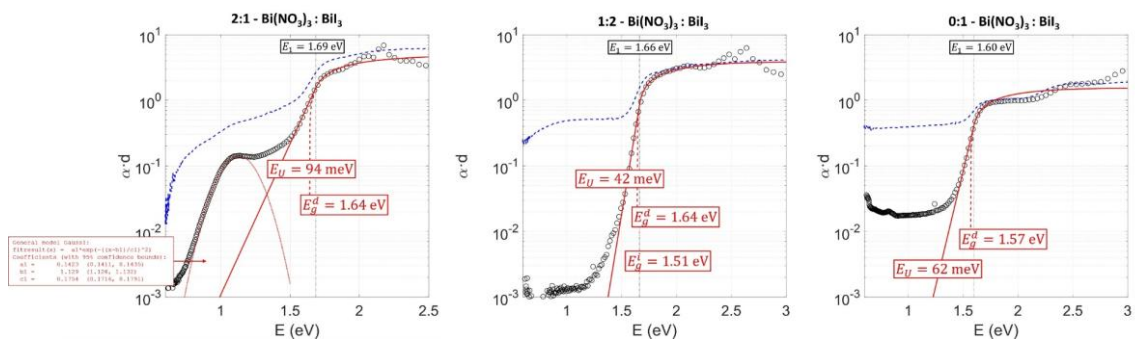

**Fig. S16.** PDS absorption for BiSI with synthesized with different content of bismuth (III) nitrate pentahydrate in the precursor molecular ink. The deposition conditions were the same for the three samples. The sub-gap absorption disappears when the content of nitrate decreases.

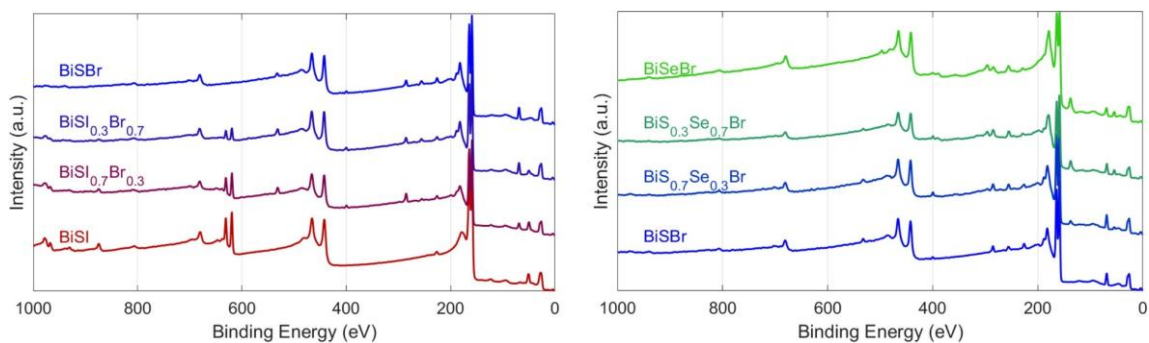

**Fig. S17.** XPS spectra for Bi-chalcohalide samples.

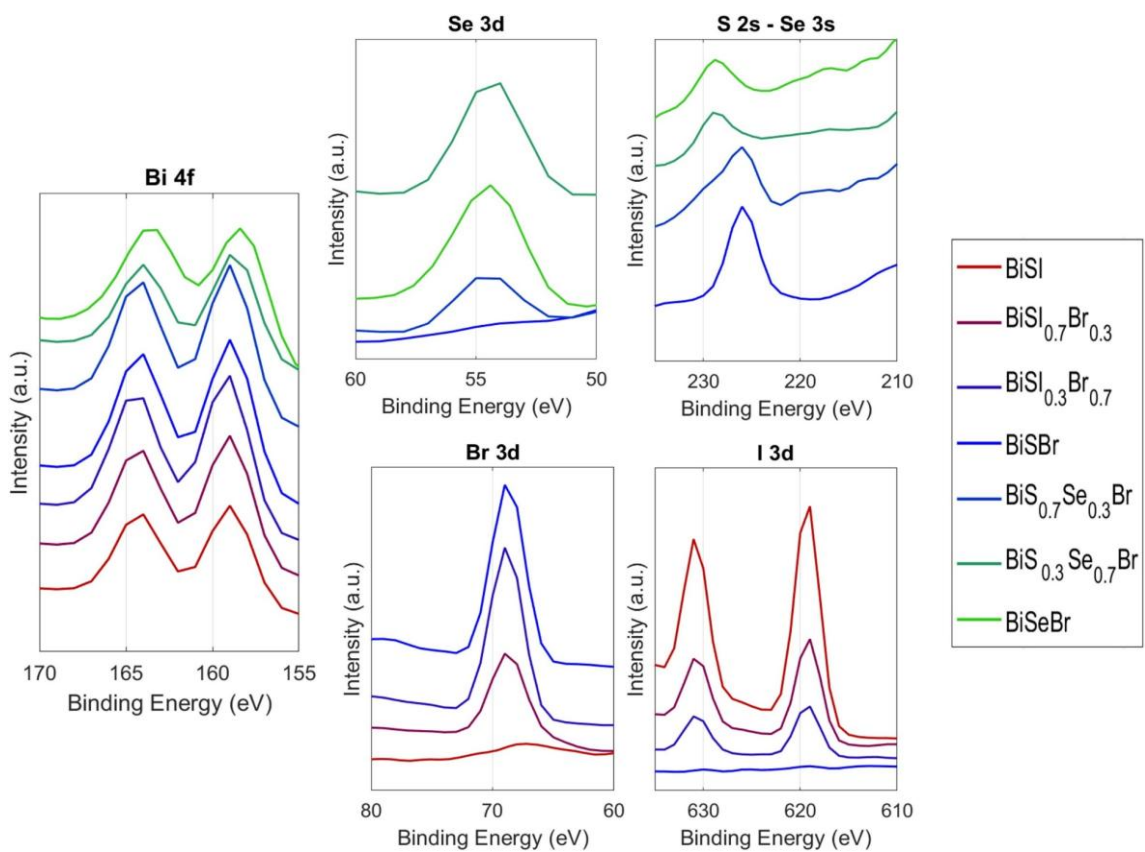

**Fig. S18.** Characteristic peaks of XPS spectra of Bi-chalcohalide solid solutions.

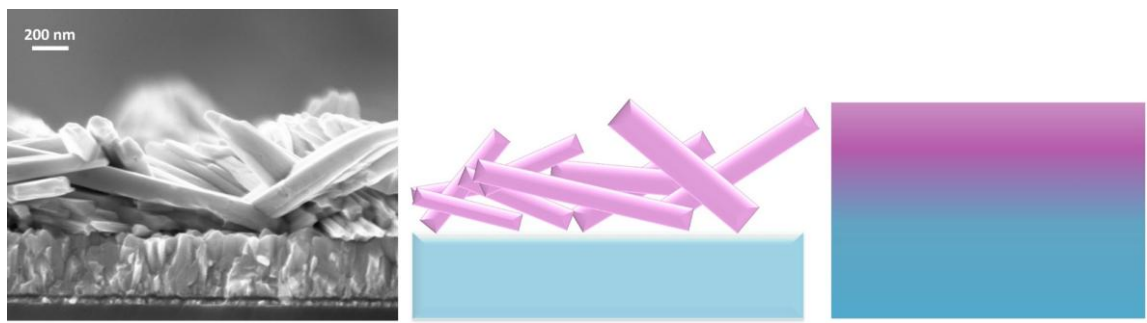

**Fig. S19.** SEM cross-sectional view of the BiSI parent compound on an FTO substrate (**left**). Schematic representation of the chalcohalide nanorod arrangement, illustrating the void spaces at the substrate interface (**centre**). Graphical depiction of the interface concentration gradient observed in EBS measurements, attributed to film porosity (**right**).